\documentclass[12pt]{article}
\usepackage{latexsym}
\usepackage{amsmath}
\usepackage{amssymb}
\usepackage{relsize}
\usepackage{geometry}

\usepackage{tensor}
\usepackage{physics} 
\usepackage{csquotes} 

\usepackage{graphicx}       	
\usepackage[font=small,labelfont=bf]{caption}
\usepackage{subcaption}

\def\beqn{\begin{eqnarray}}
\def\eeqn{\end{eqnarray}}

\def\beq{\begin{equation}}
\def\eeq{\end{equation}}
\def\ba{\beq\new\begin{array}{c}}
\def\ea{\end{array}\eeq}

\newcommand{\gsim}{\lower.7ex\hbox{$
\;\stackrel{\textstyle>}{\sim}\;$}}
\newcommand{\lsim}{\lower.7ex\hbox{$
\;\stackrel{\textstyle<}{\sim}\;$}}

\newcommand{\ntwo}{${\mathcal N}=2$ }

\newcommand{\ntwot}{${\mathcal N}= \left(2,2\right) $ }

\newcommand{\pt}{\partial}

\numberwithin{equation}{section}

\newcommand{\p}{\partial}

\def\slashed#1{\setbox0=\hbox{$#1$}             
   \dimen0=\wd0                                 
   \setbox1=\hbox{/} \dimen1=\wd1               
   \ifdim\dimen0>\dimen1                        
      \rlap{\hbox to \dimen0{\hfil/\hfil}}      
      #1                                        
   \else                                        
      \rlap{\hbox to \dimen1{\hfil$#1$\hfil}}   
      /                                         
   \fi}                                        %

\newcommand{\wcpt}{$\mathbb{WCP}(2,2)\;$}

\usepackage{mathtools}
\usepackage{hyperref}
\usepackage{xcolor}
\usepackage{tikz-cd}
\usepackage{placeins}
\usepackage[toc,page, title, titletoc]{appendix}
\begin{document}

\hypersetup{%
	linkbordercolor=blue,
}

%
%

\begin{titlepage}



\begin{center}
	\Large{{\bf 
	Critical Non-Abelian Vortex and \\
		Holography for Little String Theory 
	}}
	
\vspace{5mm}
	
{\large  \bf E.~Ievlev$^{\,a,b}$ and  A.~Yung$^{\,a,c}$}
\end{center}
\begin{center}

	$^{a}${\it National Research Center ``Kurchatov Institute'',
	Petersburg Nuclear Physics Institute, Gatchina, St. Petersburg
	188300, Russia}\\
%
$^{b}${\it Saint Petersburg State Electrotechnical University,
		ul. Professora Popova, St.~Petersburg 197376, Russia}\\
{\it  $^{c}$William I. Fine Theoretical Physics Institute,
University of Minnesota,
Minneapolis, MN 55455}\\
	
	\end{center}

\vspace{5mm}

\begin{center}
{\large\bf Abstract}
\end{center}

It has been shown that non-Abelian vortex string supported in four dimensional (4D)
${\mathcal N}=2$ supersymmetric QCD (SQCD) with the U(2) gauge group and
$N_f = 4$ quark flavors becomes a critical superstring. This string propagates in the ten dimensional space formed by a product of the flat 4D space and an internal space given by a Calabi-Yau non-compact threefold, namely, the conifold.
The spectrum  of closed string states of the associated string theory was obtained using the equivalence between the critical string on the conifold and the non-critical  string on the semi-infinite cigar described by SL($2, \mathbb{R}$)/U(1) Wess-Zumino-Novikov-Witten  model.  This spectrum was identified with the spectrum of hadrons in 4D ${\mathcal N}=2$  SQCD. In order to describe effective interactions of these 4D hadrons in this paper we study correlation functions of normalizable 
vertex operators localized near the tip of the SL($2, \mathbb{R}$)/U(1) cigar. We also compare our solitonic string approach
to the gauge-string duality to the  AdS/CFT-type holography for  little string theories (LSTs).  The latter  relates off mass-shell correlation functions on the  field theory side to correlation functions of non-normalizable vertex operators on the cigar. We show that in most channels holographic approach works in our theory because normalizable and non-normalizable vertex operators with the same conformal dimension are related due to the reflection from the tip of the cigar. However, we find that
holography does not work for lightest hadrons with  given baryonic charge.

\end{titlepage}

\newpage

\tableofcontents

\newpage

%
%

\section{Introduction} 
\label{sec:introduction}

Non-Abelian vortices were first found in 4D
\ntwo supersymmetric QCD (SQCD) with the gauge group U$(N)$ and $N_f \ge N$ flavors of quark hypermultiplets
\cite{HT1,ABEKY,SYmon,HT2}.
The non-Abelian vortex string is 1/2
BPS saturated and, therefore,  has \ntwot supersymmetry on its world sheet.
In addition to four translational moduli characteristic of the  Abrikosov-Nielsen-Olesen (ANO) strings 
\cite{ANO}, the non-Abelian string carries orientational  moduli, as well as the size moduli if $N_f>N$
\cite{HT1,ABEKY,SYmon,HT2} (see \cite{Trev,Jrev,SYrev,Trev2} for reviews).

Recently it was discovered that the non-Abelian solitonic vortex string in 4D \ntwo supersymmetric QCD at strong coupling  can become a critical superstring \cite{SYcstring}.
This was shown to happen in the theory with the U(2) gauge group, four quark flavors and the
Fayet-Iliopoulos (FI) \cite{FI} parameter $\xi$.  
Due to the extended supersymmetry,  the gauge coupling  in the 4D SQCD could be renormalized only at one loop.
With the judicial choice of the matter sector the one-loop renormalization cancels. No dynamical scale parameter 
$\Lambda$ is generated in  4D SQCD \footnote{However, conformal invariance of 4D SQCD is broken by the Fayet-Iliopoulos term.}.

The dynamics of the internal (orientational and size) moduli of the non-Abelian vortex string in the case at hand ($N=2$, $N_f=4$) is described by the so-called two dimensional (2D) weighted $\mathbb{CP}$ sigma model, which we denote as $\mathbb{WCP}(2,2)$, see Sec.~\ref{sec:wcp} below. Its $\beta$ function vanishes,  and the 
dimension of the  target space of \wcpt model is equal to six. \cite{SYcstring}.  Thus, in this case the target space of the world sheet theory of the non-Abelian vortex string is ten dimensional as required for a superstring to become critical.
It has a structure 
 $\mathbb{R}^4\times Y_6$, where $Y_6$ is a non-compact six dimensional Calabi-Yau manifold, the 
resolved conifold \cite{Candel,NVafa}. 
This allows one to apply the string theory for the calculation  of the spectrum of closed string states \cite{KSYconifold}. This spectrum was  interpreted   as a spectrum of hadrons in 4D \ntwo SQCD. The vortex string at hand was identified as the superstring theory of type IIA \cite{KSYconifold}.

A version of the string-gauge duality for 4D SQCD was proposed \cite{SYcstring}: at
weak coupling this theory is in the Higgs phase and can be described in terms
of quarks and Higgsed gauge bosons, while at strong coupling hadrons of this theory can be understood as string states formed by the non-Abelian vortex string. Below we will call this approach ''solitonic string-gauge duality''.

The study of the above vortex string from the standpoint of string theory, with the focus on massless states in four dimensions has been started   in \cite{KSYconifold,KSYcstring}. 
Generically,  most of the massless modes have  non-normalizable wave functions over the conifold $Y_6$, i.e. they are not localized in 4D 
and, hence, cannot be interpreted as dynamical states in 4D SQCD.  In particular, no massless 4D gravitons or vector fields were found  in the physical spectrum in \cite{KSYconifold}. However, a single massless BPS hypermultiplet in the 4D SQCD was detected 
at a self-dual point (at strong coupling). It is associated with deformations of a complex structure of the conifold and was  interpreted  as a composite 4D \textquote{baryon}%
\footnote{If the gauge group is U(2), as is our case, there are no {\em bona fide} baryons. We still use the term baryon because of a particular value of its  charge $Q_B$(baryon) = 2 with respect to the global unbroken U(1)$_B$, see Sec.~\ref{sec:scalar_spin2}.}%
.
Later the existence of this massless hypermultiplet was confirmed using purely field theoretical methods \cite{Ievlev:2020qch}.

The low lying massive string states were found in \cite{SYlittles}, and the corresponding multipet structure with respect to the 4D \ntwo supersymmetry was identified in \cite{SYlittmult}.
To analyze the massive states, a different approach was applied,
similar to the one  used for Little String Theories (LSTs), see  \cite{Kutasov} for a review.
This is the equivalence between the 
critical string  on the conifold and non-critical $c=1$ string which contains the Liouville 
field and a compact scalar at 
the self-dual radius \cite{GVafa,GivKut}. Generically the above  equivalence is formulated between the critical string on non-compact Calabi-Yau spaces with 
isolated singularity on the one hand, and non-critical $c=1$ string with the additional Ginzburg-Landau
\ntwo superconformal system \cite{GivKut} on the other hand. In  the conifold case  this extra Ginzburg-Landau CFT 
 is absent \cite{GivKutP}.

The above $c=1$ Liouville  theory has a mirror description \cite{HoriKapustin}. In this formulation it is given by supersymmetric  version of Witten's two-dimensional
black hole with a semi-infinite cigar target space \cite{Wbh}, which is the ${\rm SL}(2,R)/{\rm U}(1)$ coset Wess-Zumino-Novikov-Witten (WZNW) theory 
\cite{GVafa,GivKut,MukVafa,OoguriVafa95}.  It can be    analyzed by virtue of algebraic methods  and 
the spectrum of primary operators was computed exactly \cite{MukVafa,DixonPeskinLy,Petrop,Hwang,EGPerry}.
These exact results were used in \cite{SYlittles} to obtain a low lying spectrum of hadrons in \ntwo 4D SQCD. In the string description 4D SQCD hadrons  are associated with normalizable vertex operators localized near the tip of the cigar.

In this paper we make a next step and apply string theory description to study interactions of 4D hadrons. To this end we calculate  correlation functions of normalizable vertex operators in WZNW ${\rm SL}(2,R)/{\rm U}(1)$ coset model.
These correlation functions  were studied by many authors (see e.g. \cite{Dorn:1994xn,Zamolodchikov:1995aa,Teschner:1997ft,Teschner:1999ug,Dijkgraaf:1991ba} and a review \cite{Nakayama:2004vk}) and we apply already known results to our case.

Another purpose of this paper is to compare the solitonic string-gauge duality to AdS/CFT approach. The former identify closed string states of the string theory for the solitonic non-Abelian vortex with hadrons of 4D \ntwo SQCD.  The latter developed for much broader class of theories relates the open string description of a field theory on the world volume of $N_b$ parallel $D$-branes and observables in a theory of a closed string propagating in the background of these branes, see for example \cite{AharGubserMaldaOoguriOz} for a review. We summarize the distinctions of two approaches and suggest that our solitonic string-gauge duality can be thought of as a ''no branes''  limit of AdS/CFT-type correspondence, $N_b\to 0$ \footnote{The validity of our approximation is not related to the large $N_b$ limit, see Sec. 3.1} for the cigar background. However,   holography seems to be a distinctive feature of AdS/CFT correspondence not present in the solitonic string-gauge duality. 

To clarify this issue  we compare our solitonic string-gauge duality to the  AdS/CFT-type holography for LSTs. In 
\cite{ABKS,GivKut,GivKutP} it was argued that non-critical string theories with the dilaton linear in the Liouville coordinate $\phi$  are holographic. 
The main example of this behavior is non-gravitational LST
in the flat six-dimensional space formed by the world volume of $N_b$ parallel NS5 branes.
The string coupling exponentially 
goes to zero in the bulk of the  space-time ( at large $\phi$ )   and non-trivial dynamics (LST)  is localized near the branes.
Much in the same way as in the AdS/CFT holography off mass-shell correlation functions on the field theory side (in LST) 
correspond to string theory correlation functions on the ''boundary'' at $\phi\to\infty$.  More precisely off-shell
correlation functions in LST corresponds to correlation functions of non-normalizable vertex operators on the cigar, see \cite{Kutasov} for a review.

In this paper we test  this holography for the  string theory of our non-Abelian vortex. 
 We find that in most channels holographic approach works  because normalizable and non-normalizable vertex operators with the same conformal dimension are related due to the reflection from the tip of the cigar. 

This relation was already established in 
 \cite{LSZinLST}  in the context of the six dimensional LST in the background of parallel NS5 branes. Near the pole associated with
a physical state with mass $M$ the non-normalizable vertex operator $V^{{\rm non-norm}}$ behaves as
\beq
V^{{\rm non-norm}} \sim \frac1{p_{\mu}^2 + M^2}\, V^{{\rm norm}},
\label{LSZ}
\eeq
where $p_{\mu}$ is the momentum of the field theory state (4D momentum in our case), while $V^{{\rm norm}}$ is the normalizable vertex operator of the state with
mass $M$. Authors of  \cite{LSZinLST} call these poles as LSZ poles.

We test the holography relation \eqref{LSZ} in our theory. First, we confirm that LSZ poles of  two point correlation functions of non-normalizable vertex operators match the mass spectrum found previously \cite{SYlittles,SYlittmult}. Next we consider three point correlation functions and 
confirm the holography relation \eqref{LSZ}. The only exceptions are correlation functions of vertex operators which are on the borderline
between normalizable and non-normalizable states. These operators correspond to logarithmically normalizable wave functions in our theory and are associated with physical states in 4D SQCD, namely lightest baryons with given baryon charge. Correlation functions of these operators give scattering amplitudes of  above mentioned  baryons. Although these operators have non-normalizable ''partners'' with  same conformal dimensions, correlation functions of latter operators does not have required  LSZ pole structure.  

The paper is organized as follows. 
In Sec.~\ref{sec:nastrings} we review the non-Abelian string, and its relation to the critical superstring on the conifold.
Next, in Sec.~\ref{sec:c=1} we review the non-critical $c=1$ string theory and the spectrum of stringy hadrons in 4D SQCD. In Sec. 4 we discuss general features of our solitonic string-gauge duality versus holographic dualities.
In Sec.~\ref{sec:operators_2pt} we consider two point correlation functions for operators corresponding to baryons of the underlying 4D \ntwo SQCD. 
Three point correlation functions for such operators are studied in Sec.~\ref{sec:3pt_VVV}.
We consider these correlation functions with normalizable as well as non-normalizable operators and test the holography relation.
In Sec.~\ref{sec:cont_rep} we study correlation functions of operators from continuous representations and discuss an interpretation of such states in terms of the 4D \ntwo SQCD.
Sec.~\ref{sec:conclusions} presents our conclusions.
In Appendix~\ref{sec:3pt_additional} we cite the three point function formula in the supersymmetric Liouville theory and apply it for the case of the continuous representations.
In Appendix~\ref{sec:nosusy} we make a comparison of the supersymmetric and non-supersymmetric Liouville theories, present an idea of factorization and apply it to the supersymmetric Liouville theory.
In Appendix~\ref{sec:2pt_check} we make a consistency check of the two point correlation function formula. 
Appendix~\ref{sec:useful} contains some useful formulae.

%
%

\section {Non-Abelian vortices}
\label{sec:nastrings}
\setcounter{equation}{0}

\subsection{Four-dimensional \boldmath{${\mathcal N}=2\;$} SQCD}

As was already mentioned above non-Abelian vortex-strings were first found in 4D
\ntwo SQCD with the gauge group U$(N)$ and $N_f \ge N$ quark flavors (i.e.  hypermultiplets)
supplemented by the FI $D$ term $\xi$
\cite{HT1,ABEKY,SYmon,HT2}, see for example \cite{SYrev} for a detailed review of this theory.
Here we just mention that at weak coupling, $g^2\ll 1$, this theory is in the Higgs phase in which the scalar
components of the quark multiplets (squarks) develop vacuum expectation values (VEVs). These VEVs breaks 
the U$(N)$ gauge group
Higgsing  all gauge bosons. The Higgsed gauge bosons combine with the screened quarks to form long \ntwo multiplets with mass $m \sim g\sqrt{\xi}$.

Since the \ntwo SQCD is in the Higgs phase, non-Abelian 
vortex strings confine monopoles. In the \ntwo 4D theory these strings are 1/2 BPS-saturated; hence,  their
tension  is determined  exactly by the FI parameter,
\beq
T=2\pi \xi\,.
\label{ten}
\eeq
However, non-Abelian vortices in U(N) theories are topologically stable and cannot
be broken. Therefore the finite-length strings are closed.
In particular, the monopoles cannot be attached to the string endpoints. In fact, in the U$(N)$ theories confined  
 monopoles 
are  junctions of two distinct elementary non-Abelian strings \cite{SYmon,HT2,T} (see \cite{SYrev} 
for a review). As a result,
in  four-dimensional \ntwo SQCD we have 
monopole-antimonopole mesons in which the monopole and antimonopole are connected by two confining strings.
 In addition, in the U$(N)$  gauge theory we can have baryons  appearing as  a closed 
\textquote{necklace} configurations of $N\times$(integer) monopoles \cite{SYrev}.

Squark VEVs lead to the color-flavor locking phenomenon: the global flavor symmetry group SU$(N_f)$ and the gauge group U($N$) are broken down, and the resulting global symmetry is
\beq
 {\rm SU}(N)_{C+F}\times {\rm SU}(N_f-N)\times {\rm U}(1)_B,
\label{c+f}
\eeq
see \cite{SYrev} for more details. 

The unbroken global U(1)$_B$ factor above is identified with a baryonic symmetry. Note that 
what is usually identified as the baryonic U(1) charge is a part of  our 4D theory  gauge group.
 ``Our" U(1)$_B$
is  an unbroken by squark VEVs combination of two U(1) symmetries:  the first is a subgroup of the flavor 
SU$(N_f)$ and the second is the global U(1) subgroup of U$(N)$ gauge symmetry.

The stringy monopole-antimonopole mesons as well as monopole baryons with spins $J\sim 1$ are massive.
Their masses are determined by the string tension,  $\sim \sqrt{\xi}$.
They are heavy at weak coupling and decay into perturbative quarks and gauge bosons, which have mass 
$m\sim g\sqrt{\xi}$. Instead at strong coupling \ntwo SQCD is in the ''instead-of-confinement'' phase 
\cite{SYdualrev,Ievlev:2020qch}  where perturbative states evolve into hadrons formed by the closed string states.
 	All hadrons found from string theory of non-Abelian vortex  as closed string states turn out to be baryons and look like monopole ``necklaces'' \cite{SYlittles}.

To conclude this section, we note that our 4D \ntwo SQCD  has a Higgs branch formed by massless quarks which are in  the 
bi-fundamental representation of the global group \eqref{c+f} and carry baryonic charge, see \cite{KSYconifold} for more details.
This Higgs branch  is hyper-K\"ahlerian \cite{SW2,APS}, therefore 
its metric cannot be 
modified by quantum corrections \cite{APS}. In particular, once the Higgs branch is present at weak coupling
we can continue it all the way into strong coupling. Thus, at strong coupling in the ''instead-of-confinement'' phase besides 
stringy hadrons we also have massless perturbative states: bi-fundamental quarks. It is argued in \cite{SYlittles} that they
play an important role in the stringy dynamics, see Sec.~\ref{sec:cont_rep}.

\subsection{World sheet sigma model}
\label{sec:wcp}

The presence of non-Abelian subgroup SU$(N)_{C+F}$ in \eqref{c+f} is the reason for the formation of the
non-Abelian vortex strings \cite{HT1,ABEKY,SYmon,HT2} in the 4D SQCD.
The most important feature of these vortices is the presence of the so-called orientational  zero modes.
In this subsection we are going to briefly review the sigma model emerging on the world sheet
of the non-Abelian critical string supported in \ntwo SQCD with $N=2$, $N_f=4$ \cite{SYcstring,KSYconifold,KSYcstring}.

If $N_f=N$  the dynamics of the orientational zero modes of the non-Abelian vortex, which become 
orientational moduli fields 
 on the world sheet, is described by two-dimensional
\ntwot supersymmetric $\mathbb{CP}(N-1)$ model \cite{SYrev}.

Upon adding extra quark flavors non-Abelian vortices become semilocal.
They acquire so-called size moduli \cite{AchVas}.  
In particular, for the non-Abelian semilocal vortex at hand with $N_f=4$,  in 
addition to  the orientational zero modes  $n^P$ ($P=1,2$), there are size moduli which we denote as   
$\rho^K$ ($K=1,2$) \cite{HT1,HT2,AchVas,SYsem,Jsem,SVY}.  

The target space of the resulting weighted $\mathbb{CP}$ sigma model ($\mathbb{WCP}(2,2)$ for short)
is a symplectic quotient $\mathbb{C}^4 // $U(1), or in other words the U(1) $D$-term condition
\begin{equation}
	|n^P|^2-|\rho^K|^2 = \beta\,.
\label{Dterm}
\end{equation}
plus U(1) gauge invariance. This quotient has complex dimension three.
 
Apart from these internal modes, the string under consideration has of course the usual translational modes corresponding to the ${\mathbb R}^4$ space that the string lives in.
In total, the number of real bosonic degrees of freedom in this model is ten (four translational plus six internal). 
These degrees of freedom form a ten-dimensional space needed 
for superstring to be critical.

Since non-Abelian  vortex string  is 1/2 BPS it preserves ${\mathcal N} =(2,2)$  in the world sheet
sigma model necessary for the \ntwo space-time supersymmetry  \cite{Gepner,BDFM}. Moreover, as was shown in \cite{KSYconifold},
 the string theory of the non-Abelian critical vortex is of type IIA.

The world sheet coupling constant $\beta$  in (\ref{Dterm}) is related to the bulk gauge coupling constant $g^2$.
At weak coupling the relation is given by \cite{SYrev}
\beq
\beta\approx \frac{4\pi}{g^2}\,.
\label{betag}
\eeq
Note that the first (and the only) coefficient of the beta-function is the same for the 4D SQCD and the
world-sheet model. Both vanish at $N_f=2N$. This ensures that our world-sheet theory is conformal.
Therefore,
its target space is Ricci flat and (being K¨ahler due to \ntwot supersymmetry)
represents a non-compact Calabi-Yau manifold, namely the conifold $Y_6$, see
\cite{NVafa} for a review.

The global symmetry of the world-sheet sigma model is
\beq
 {\rm SU}(2)\times {\rm SU}(2)\times {\rm U}(1)\,,
\label{globgroup}
\eeq
i.e. exactly the same as the unbroken global group in the 4D theory, cf. (\ref{c+f}), at  $N=2$ and $N_f=4$. 
The fields $n$ and $\rho$ 
transform in the following representations:
\beq
n:\quad (\textbf{2},\,0,\, 0), \qquad \rho:\quad (0,\,\textbf{2},\, 1)\,.
\label{repsnrho}
\eeq

\subsection{Thin string regime}
\label{thinstring}

The coupling constant of 4D SQCD can be complexified
\beq
 \tau \equiv i\frac{4\pi}{g^2} +\frac{\theta_{4D}}{2\pi}\,,
\label{bulkduality}
\eeq 
 where $\theta_{4D}$ is the four-dimensional $\theta$ angle.
Note that SU$(N)$ version of the four-dimensional SQCD at hand 
possesses a strong-weak coupling ($S$) duality, namely, $\tau\to  -\frac{1}{\tau}\,$ \cite{APS,ArgPlessShapiro}.
There is also a symmetry with respect to the shift $\theta_{4D} \to \theta_{4D} + 2\pi$ ($T$-duality).

The two-dimensional coupling constant $\beta$ can be naturally complexified too if we include the two-dimensional
$\theta$ term,
\begin{equation}
	\beta \to \beta + i\,\frac{\theta_{2D}}{2\pi}\,.
\end{equation}

The exact relation between the complexified 4D and 2D couplings was derived in \cite{Ievlev:2020qch}:
\begin{equation}
	e^{- 2 \pi \beta} = \lambda(2 \tau + 1),
	\label{taubeta}
\end{equation}
where $\lambda$ is the elliptic modular lambda function. It can be expressed in terms of standard Jacobi $\theta$-functions 
\cite[p. 108]{Chandrasekharan} as 
\begin{equation}
	\lambda(\tau) 
		= \frac{\theta_1^4(\tau) }{\theta_3^4(\tau) } \,.
\label{lambda_function}		
\end{equation}
Equation (\ref{taubeta}) generalizes the quasiclassical relation
\eqref{betag}. It was derived using 2D-4D correspondence, namely, the match of the BPS spectra of the 4D theory 
at $\xi=0$ and the world-sheet theory on the non-Abelian string \cite{SYmon,HT2,Dorey,DoHoTo}. 
Note that our result \eqref{taubeta} differs from that
obtained in \cite{Karasik} by the shift of $\theta_{4D}$ by $\pi$.

According to the hypothesis formulated in  \cite{SYcstring}, our critical non-Abelian string becomes infinitely thin at  strong coupling at the critical  point $\tau_c$.  Moreover, in
\cite{KSYconifold} it was conjectured  that $\tau_c$ corresponds to $\beta =0$ in the world-sheet theory via relation \eqref{taubeta}. The vortex transverse size is given by $1/m$, where $m$ is the mass of 4D SQCD perturbative states.
Thus, we assume that the mass of the 4D perturbative states $m \to \infty$ at $\beta=0$, which corresponds to $\tau = \tau_c$ in 4D SQCD.
 
At the    point $\beta=0$  
the non-Abelian string 
becomes infinitely thin,  higher derivative terms can be neglected and the world sheet theory of the non-Abelian 
string reduces to the $\mathbb{WCP}(2,2)$ model. The point $\beta=0$ is a natural choice because at this point
 we have a regime change in the 2D sigma model. 
This is the point where the {\em resolved} conifold defined by the $D$ term constraint
(\ref{Dterm}) develops a conical singularity \cite{NVafa}. Recently it was shown in \cite{Ievlev:2020qch} that the point 
$\tau_c$, where our non-Abelian vortex string becomes thin is $\tau_c = 1$, which corresponds to $\beta=0$ via 
\eqref{taubeta} \footnote{This corrects the value $\tau = i$ suggested earlier in \cite{KSYconifold,KSYcstring}.}. This point is self-dual under a combination of $S$ and $T$-dualities. 

When the parameter $\beta$ vanishes, the conifold singularity can be smoothened by a complex structure {\em deformation}. 
It was shown in \cite{KSYconifold} that the complex parameter $b$ of this  deformation corresponds to a massless particle - a baryon with the baryon charge $Q_B = 2$.
This baryon can be represented as a \textquote{necklace} of four monopoles \cite{Ievlev:2020qch}.

%
%

\section{Non-critical \boldmath{$c=1$} string }
\label{sec:c=1}

As was mentioned in Sec.~\ref{sec:introduction}, the spectrum of massive string modes on the conifold  was found in 
\cite{SYlittles} ( see also \cite{SYlittmult}) using an alternative formulation in terms of the $c=1$ string theory.
This theory contains the Liouville field and a compact scalar at 
the self-dual radius and is argued to be equivalent   to the string theory on the conifold in the so called double scaling limit \cite{GVafa,GivKut,GivKutP}.
In this section we briefly review this theory and spectrum of physical string states obtained in \cite{SYlittles}.

The authors of \cite{GivKut,GivKutP} considered the above alternative formulation as a kind of AdS/CFT-type duality \footnote{Below in this paper we clarify the relation of our approach to AdS/CFT-type holography.}.
Here we follow the logic of Ref.~\cite{GVafa} \footnote{In this paper the equivalence was shown for topological versions of two string theories.} and consider it as an equivalence of two string theories: the critical string on the conifold  and non-critical  $c=1$ string with Liouville field, see also discussion in Sec.~\ref{sec:hologr_views} for more details.

\subsection{Liouville theory}

Relevant non-critical $c=1$ string theory is formulated on the target space
\beq
\mathbb{R}^4\times \mathbb{R}_{\phi}\times S^1,
\label{target}
\eeq
where $\mathbb{R}_{\phi}$ is a real line associated with the Liouville field $\phi$ and the theory  
has a linear in $\phi$ dilaton, such that string coupling is given by
\beq
g_s =e^{-\frac{Q}{2}\phi}\, .
\label{strcoupling}
\eeq

In \cite{GivKut} the authors proposed a more general  equivalence between the critical string on non-compact Calabi-Yau spaces with 
isolated singularity on the one hand, and non-critical $c=1$ string with the additional Ginzburg-Landau
\ntwo superconformal system on the other hand. 
In our conifold case  this extra Ginzburg-Landau factor  is absent from \eqref{target}, see \cite{GivKutP}.

The bosonic stress tensor of the $c=1$ matter coupled to 
2D gravity is given by (cf. the linear dilaton \eqref{strcoupling})
\beq
T_{--}= -\frac12\,\left[(\pt_z \phi)^2 + Q\, \pt_z^2 \phi + (\pt_z Y)^2\right]. 
\label{T--}
\eeq
The  compact scalar $Y$  represents $c=1$ matter and satisfies the periodicity condition
\begin{equation}
	Y \sim Y+2\pi Q\, .
\label{Y_period}
\end{equation}
The scalars are normalized so that their propagators are
\beq
\langle \phi(z),\phi(0)\rangle = -\log{z\bar{z}}, \qquad \langle Y(z),Y(0)\rangle = -\log{z\bar{z}}\,.
\label{propagators}
\eeq
The central charge of the supersymmetric version of this $c=1$ theory is
\beq
c_{\phi+Y}^{SUSY} = 3 + 3Q^2.
\label{cphiY}
\eeq
In order for the string on \eqref{target} to be critical this central charge should be 
equal to 9. This gives 
\beq
Q=\sqrt{2}\,.
\label{Q}
\eeq

Deformation of the conifold (complex structure deformation) mentioned above translates into adding the Liouville interaction
to the world-sheet sigma model \cite{GivKut}
\beq
\delta L= b\int d^2\theta \, e^{-\frac{\phi + iY}{Q}}\,.
\label{liouville}
\eeq
Here $b$ is the deformation parameter.
The conifold singularity at  $b=0$ corresponds to the string coupling constant becoming infinitely large at
$\phi \to -\infty$, see \eqref{strcoupling}. At $b\neq 0$ the Liouville interaction regularize the behavior
of the string coupling preventing the string from propagating to the region of large negative $\phi$.

The mirror description of the Liouville $c=1$ non-critical string theory \cite{HoriKapustin} is in terms of two-dimensional
black hole \cite{Wbh}, which is the SL($2, \mathbb{R}$)/U(1) coset WZNW theory 
\cite{GVafa,GivKut,MukVafa,OoguriVafa95} 
at the level
\begin{equation}
	k =\frac{2}{Q^2}\,.
\label{k_Q_relation}
\end{equation}
In the case of the conifold ($Q=\sqrt{2}$) this gives
\beq
k=1,
\label{k=1}
\eeq 
where $k$ is the total level of the Ka\v{c}-Moody algebra in the supersymmetric version (the level
of the bosonic part of the algebra is then $k_b=k+2 =3$). The target 
space of 
this theory has the form of a semi-infinite cigar;  the field $\phi$ associated with the motion along the 
cigar
cannot take large negative values due to semi-infinite geometry. 

In this description the string
coupling constant is largest at the tip of the cigar, $g_s \sim 1/b$. We assume that parameter $b$ is large so the string
coupling at the tip of the cigar is small  and the string perturbation theory becomes reliable. In particular,
we can use the tree-level approximation to obtain the spring spectrum. Note also that, as we already mentioned
in the Introduction, the SL($2, \mathbb{R}$)/U(1) WZNW model is exactly solvable so we do not need to keep the level of the Ka\v{c}-Moody algebra $k$ large and can go to the strong coupling at $k=1$, see \eqref{k=1}.

\subsection{Vertex operators }
\label{sec:vertices}

Vertex operators for  the  string theory on \eqref{target} are constructed in \cite{GivKut}, see also 
\cite{GivKutP,MukVafa}. Primaries of the $c=1$  part for large 
positive $\phi$ (where the target space becomes a cylinder $\mathbb{R}_{\phi}\times S^1$) take the form
\beq
V_{j;m_L,m_R}\approx \exp{\left(\sqrt{2}j\phi + i\sqrt{2}(m_L Y_L +m_R Y_R)\right)},
\label{vertex}
\eeq
where we split  $Y$ into left and right-moving parts.
 For the self-dual radius \eqref{Q} (or $k=1$) the parameter $2m$ in Eq. (\ref{vertex}) 
is integer. For the
left-moving sector $2m_{L}\equiv 2m$ is 
the total momentum plus the winding number along the compact dimension $Y$. For the right-moving sector
we introduce $2m_{R}$ which is the momentum minus the winding number.  For our case type IIA string $m_{R}= - m$, while for type IIB string $m_{R}= m$ \cite{SYlittmult}.

The primary operator \eqref{vertex} is related  to the wave 
function over ``extra dimensions'' as follows:
$$V_{j;m_L,m_R} = g_s \Psi_{j;m_L,m_R}(\phi,Y)\,.$$ The string coupling 
\eqref{strcoupling}  depends on $\phi$. Thus, 
\beq
\Psi_{j;m_L,m_R}(\phi,Y) \sim e^{\sqrt{2}(j+\frac{1}{2})\phi + i\sqrt{2}(m_L Y_L +m_R Y_R)}\,.
\eeq
We will look for string states with normalizable wave functions over the ``extra dimensions'' which we will 
interpret as hadrons in 4D \ntwo SQCD. The 
condition for the string states to have  normalizable wave functions reduces to 
\beq
j\le -\frac12\,.
\label{normalizable}
\eeq
The scaling dimension of the primary operator \eqref{vertex} is 
\beq
\Delta_{j,m} = \frac1{k}\left\{m^2 - j(j+1)\right\}|_{k=1} = m^2 - j(j+1) \, .
\label{dimV}
\eeq
We include  the case $j=-\frac12$
which is at the borderline between normalizable and non-normalizable states. In \cite{SYlittles} it is shown 
that  $j=-\frac12$  corresponds to the norm logarithmically divergent in the infrared in terms of the radial coordinate of the conifold. In particular, the massless baryon $b$ belongs to states with $j=-\frac12$, see next subsection. 

Unitarity implies that the conformal dimension \eqref{dimV} should be positive,
\beq
\Delta_{j,m}> 0\,.
\label{Deltapositive}
\eeq
Moreover, to ensure that conformal dimensions of left and right-moving parts of the vertex operator
\eqref{vertex} are the same we impose that $m_R=\pm m_L$.

The spectrum of the allowed values of $j$ and $m$ in \eqref{vertex} was  exactly determined by using the Ka\v{c}-Moody algebra
for the coset ${\rm SL}(2,R)/{\rm U}(1)$ in \cite{MukVafa,DixonPeskinLy,Petrop,Hwang,EGPerry}, 
see \cite{EGPerry-rev} for a review. Both discrete and continuous representations were found. Parameters $j$
and $m$ determine the global quadratic Casimir operator and the projection of the spin  on the third axis,
\beq
J^2\, |j,m\rangle\, = -j(j+1)\,|j,m\rangle, \qquad J^3\,|j,m\rangle\, =m \,|j,m\rangle
\eeq
where $J^a$  $(a=1,2,3)$ are the global ${\rm SL}(2,R)$ currents. 

We have 

(i) {\em Discrete representations} with
\beq
j=-\frac12, -1, -\frac32,..., \qquad m=\pm\{j, j-1,j-2,...\}.
\label{discrete}
\eeq

(ii) {\em Principal} continuous representations with
\beq
j=-\frac12 +is, \qquad m= {\rm integer} \quad {\rm or} \quad m= \mbox{ half-integer},
\label{principal}
\eeq
where $s$ is a real parameter and

(iii) {\em Exceptional} continuous representations with
\beq
-\frac12 \le j < 0, \qquad m= {\rm integer}.
\label{exceptional}
\eeq

We see that discrete representations include the normalizable states localized near the tip of the cigar,
 while the continuous representations contain
non-normalizable states,\footnote{We will discuss the case $j=-\frac12$
which is on the borderline between normalizable and non-normalizable states in the next subsection.}  see \eqref{normalizable}. This nicely matches our qualitative expectations.

Discrete representations contain
states with negative norm. To exclude the ghost states a restriction for spin $j$ is imposed 
\cite{DixonPeskinLy,Petrop,Hwang,EGPerry,EGPerry-rev} 
\beq
 -\frac{k+2}{2}< j <0\,.
\label{no_ghosts}
\eeq
Thus, for  our value $k=1$ we are left with only two allowed values of $j$,
\beq
j=-\frac12, \qquad m= \pm\left\{\,\frac12,\, \frac32,...\right\}
\label{j=-1/2}
\eeq
and 
\beq
j=-1, \qquad m= \pm\{\,1, \,2,...\}.
\label{j=-1}
\eeq

\subsection{Scalar and spin-2 states} 
\label{sec:scalar_spin2}

Four-dimensional spin-0 and spin-2 states were found in \cite{SYlittles} using vertex operators \eqref{vertex}.
The 4D scalar vertices $V^S$   in the $(-1,-1)$ picture  have the
form \cite{GivKut}
\beq
V^{S}_{j;m,-m}= e^{-\varphi_L -\varphi_R }\, e^{ip_{\mu}x^{\mu}}\,
 V_{j;m,-m}\, ,
\label{tachyon}
\eeq
where superscript $S$ stands for scalar, $\varphi_{L,R}$ represent bosonised ghosts in the left and right-moving sectors, while  $p_{\mu}$ is the 4D momentum of the string state.

The condition for the  state \eqref{tachyon} to be physical is
\beq
\frac12 +\frac{p_{\mu}p^{\mu}}{8\pi T} + m^2 - j(j+1) = 1,
\label{tachphys}
\eeq
where \eqref{dimV} was used. The first term on the l.h.s. comes from ghosts 
(recall that the conformal dimension
of the ghost operator $e^{q\varphi}$ is equal to $-(q+q^2/2)$ in the picture $q$).

The GSO projection restricts the integer $2m$ for the operator in \eqref{tachyon} to be odd
\cite{KutSeib,GivKut},
\beq
 m=\frac12 +\mathbb{Z}.
\label{oddn}
\eeq
For half-integer $m$ we have only one possibility $j=-\frac12$, see \eqref{j=-1/2}.
This determines the masses of the 4D scalars, 
\beq
\frac{(M^S_{m})^2}{8\pi T}=-\frac{p_{\mu}p^{\mu}}{8\pi T} = m^2 -\frac14 \,,
\label{tachyonmass}
\eeq
where the Minkowski 4D metric of signature $(-1,1,1,1)$ is used.
It is shown in \cite{SYlittles} that the state with $m=\pm 1/2$ is the massless baryon $b$, associated with deformations of the conifold complex structure \cite{KSYconifold}. It consists of four monopoles in a \textquote{necklace}  formed by a closed string \cite{Ievlev:2020qch}.
Higher states with $m = \pm (3/2, 5/2 ,...)$ are massive 4D scalars.

At the next level we consider  4D spin-2 states. 
The corresponding vertex operators are given by
\beq
\left(V_{j;m,-m}(p_{\mu})\right)^{{\rm spin}-2}= 
\xi_{\mu\nu}\psi^{\mu}_L\psi^{\nu}_R\,e^{-\varphi_L -\varphi_R }\, 
e^{ip_{\mu}x^{\mu}}\, V_{j;m,-m}\, ,
\label{graviton}
\eeq
where $\psi^{\mu}_{L,R}$ are the world-sheet superpartners to 4D coordinates $x^{\mu}$, while
$\xi_{\mu\nu}$ is the polarization tensor.
For these states to be physical one should impose a condition 
\beq
\frac{p_{\mu}p^{\mu}}{8\pi T} + m^2 - j(j+1) = 0\,.
\label{gravphys}
\eeq 

The GSO projection selects now $2m$ to be even, $|m|=0, 1,2,...$ \cite{GivKut},
thus we are left with only one allowed value of $j$, $j=-1$ in \eqref{j=-1}. Moreover, the value
$m=0$ is excluded.
This leads to the following expression  for the masses of spin-2 states:
\beq
(M^{{\rm spin}-2}_{m})^2 = 8\pi T\,m^2, \qquad |m|=1,2,... .
\label{gravitonmass}
\eeq
All spin-2 states are massive, and no massless 4D graviton appears in our theory. 
It  matches the fact that our  4D \ntwo QCD, is defined in flat space without gravity.

The momentum $m$ in the compact dimension is related to the baryonic charge.
It was shown in \cite{SYlittles,SYlittmult} that the baryon charge of the vertex operators \eqref{tachyon}, \eqref{graviton} is given by
\beq
 Q_B = 4m.
\label{m-baryon}
\eeq
All closed string states are baryons.

\section{Solitonic string-gauge duality versus holography }
\label{sec:hologr_views}

As we already mentioned in the Introduction a version of the string-gauge duality
for 4D \ntwo SQCD  with $N_f=2N=4$ was proposed in \cite{SYcstring}: at weak coupling this 
theory is in the Higgs phase and can be 
described in terms of (s)quarks and Higgsed gauge bosons, while at strong coupling hadrons of this theory 
can be understood as closed string states formed by the non-Abelian vortex string.
This duality was further explored by studying the string theory for the critical 
non-Abelian vortex in \cite{KSYconifold,KSYcstring,SYlittles}. 

We call this duality a  solitonic string-gauge  duality and would like to compare it with AdS/CFT-type holography.
Of course, there is a conceptual difference. In our approach ten dimensional  space is an artificial construction formed by 4D ``real'' space where SQCD lives and six dimensional conifold associated with orientational and size moduli of the non-Abelian vortex. Instead, in AdS/CFT correspondence the fundamental superstring  propagates in the ten dimensional space from the very beginning.

Also the scales of the string tension are dramatically different in two approaches. The scale of the squire root of the fundamental string tension is  determined  roughly speaking by the Plank scale, while for the solitonic non-Abelian vortex $\sqrt{T}$ is fixed at the SQCD scale set by the FI parameter, see \eqref{ten}.

However, we can take a more pragmatic point of view  ''forgetting'' for a minute where the ten dimensional space come from  and think of  our solitonic string-gauge duality as a duality which associates a given 4D SQCD with a  string theory on the conifold. Then we can summarize the main distinctions of our approach as follows.
\begin{itemize}

\item AdS/CFT correspondence is typically  based on the presence of  $N_b$ parallel $D$-branes, see for example \cite{AharGubserMaldaOoguriOz} for a review. Moreover, the  validity of the gravity approximation requires  a large $N_b$ limit. In our string theory on the conifold there are no branes. The validity of our approximation is based on the large value of the parameter of the conifold complex structure deformation $b$. Large $b$ ensures the validity of the gravity approximation because the curvature of the conifold remains everywhere small, see for example \cite{YNSflux}.
Moreover, as we already mentioned large $b$ ensures small $g_s$ in the string theory  formulated on the cigar.  

\item AdS/CFT correspondence assumes holography. Off-shell correlation functions on the field theory side  
correspond to string theory correlation functions on the ''boundary'', infinitely  far away from the branes.
Instead, in our solitonic string-gauge duality  all non-trivial ``real" physics should be localized exclusively  near the tip of the cigar. We consider only normalizable vertex operators with $j\le -\frac12$ (see 
\eqref{normalizable} and identify them with hadrons of 4D SQCD.

\end{itemize}

The first distinction above suggests that  we can think of our solitonic string-gauge duality as of $N_b=0$ limit of AdS/CFT correspondence. The simplest example is the Klebanov-Witten's construction \cite{KlebWitten} of $N_b$ $D$3-branes filling the $\mathbb{R}^4$ space near the conifold singularity in type IIB superstring. The presence of the 5-form flux sourced by branes makes a direct product of  $\mathbb{R}^4$ and $Y_6$ a warped product with AdS$_5$ geometry.  In the limit $N_b=0$ the warped factor disappears and we get our background $\mathbb{R}^4\times Y_6$.

However, the second distinction seems to be a crucial one. To clarify this issue below in this paper we compare our theory with the most close example of the AdS/CFT holography: non-gravitational six dimensional LST on the world volume of $k$ $NS$5-branes 
\cite{GivKut,ABKS,LSZinLST}.  This theory is holographically dual to the non-critical  string theory on $\mathbb{R}^6 \times 
 {\rm SL}(2,\mathbb{R})_k/{\rm U}(1)  \times  {\rm SU(2)}_k/{\rm U(1)}$,  where $k$ is the level of the Kac-Moody algebra of  WZNW model \footnote{In our string theory the last factor  (compact SU(2)/U(1) CFT) is absent.}. The duality ensures that off-shell
correlation functions in LST corresponds to correlation functions of non-normalizable
vertex operators on the cigar, see \cite{Kutasov} for a review. 

Below we test this type of holography for the string theory of our non-Abelian vortex. In particular, we test the LSZ relation \eqref{LSZ} found in \cite{LSZinLST} for the theory on $NS$5-branes. First, we find the poles of  two point correlation functions 
of non-normalizable vertex operators and compare results to the hadron spectrum found previously \cite{SYlittles,SYlittmult},
see \eqref{tachyonmass} and \eqref{gravitonmass}. The poles must directly correspond to hadron masses.
Second, we consider the correlation functions of normalizable vs. non-normalizable operators and see how they are related to each other.

To conclude this section, we note that in \cite{GivKut,GivKutP} the above holographic duality for LST on $NS$5-branes was generalized to the correspondence  between the critical string on non-compact Calabi-Yau spaces with isolated singularity in the double scaling limit on the one hand, and non-critical $c = 1$ string with the additional Ginzburg-Landau \ntwo superconformal
system on the other hand. The equivalence of the critical string on the conifold and non-critical $c = 1$ string we use here is a particular  case of this correspondence. However, this correspondence looks more like equivalence of two closed string theories rather then duality between open and closed string description typical for AdS/CFT-type holography. Therefore,
as we already mentioned  we follow the logic of Ref.~\cite{GVafa}  and consider this correspondence as an
equivalence of two string theories rather than the AdS/CFT-type duality. One of the important directions of the future work is to demonstrate this equivalence  more directly.

%
%

\section{Two point correlation functions and (non-)normalizable operators}
\label{sec:operators_2pt}

\subsection{Reflection property}
\label{sec:refl}

As we already mentioned  SL($2,\mathbb{R}$)/U(1) primary fields were constructed in \cite{GivKut}, see also \cite{GivKutP,MukVafa} 
and \cite{Teschner:1997ft,Teschner:1999ug}. 
Primaries of the $c=1$  part can be expanded  at $\phi \to \infty$ where the target space becomes a cylinder $\mathbb{R}_{\phi}\times S^1$. In \eqref{vertex} we presented the leading at large $\phi$ term for values of $j$ associated with discrete series \eqref{discrete}. For generic $j$ the asymptotic expansion can be written as follows
\footnote{ Similar formulas have appeared also in \cite{LSZinLST}, but in this paper we use the normalization of \cite{Teschner:1999ug}, see eq. (13)-(14), (16)-(20) in the latter paper.
Also, our notation for $m_R, m_L$-modes is related to the notation $m, \bar{m}$ of \cite{GivKut,LSZinLST} as $m = m_L$, $\bar{m} = - m_R$.
}
 \cite{Teschner:1999ug}, 
\begin{equation}
	V_{j, m_L, m_R}
	=
	e^{i Q(m_L Y_L + m_R Y_R)}\left[e^{Q j \phi}+ R(j, m_L, m_R ; k) e^{-Q(j+1) \phi}+\cdots\right]
\label{GK_decomposition}
\end{equation}
%
%
%
%
%
%
where the ellipses stand for terms that are subleading at large positive $\phi$.
The background charge $Q$ is related to the level $k$ via \eqref{k_Q_relation} as $k = 2/Q^2$. We keep $k$ arbitrary
and put it to the value  $k=1$ (see \eqref{k=1}) at the last step.
The so-called reflection coefficient $R(j, m_L, m_R ; k)$ is given by \cite{LSZinLST}
\begin{multline}
	R(j, m_L, m_R ; k) \\
	=
		\left[ \frac{1}{\pi} \frac{\Gamma\left(1+\frac{1}{k}\right)}{\Gamma\left(1-\frac{1}{k}\right)} \right]^{2j+1} 
		\frac{\Gamma\left(1-\frac{2 j+1}{k}\right) \Gamma(m_L+j+1) \Gamma(m_R+j+1) \Gamma(-2 j-1)}%
		{\Gamma\left(1+\frac{2 j+1}{k}\right) \Gamma(m_L-j) \Gamma(m_R-j) \Gamma(2 j+1)}
\label{refl_GK}
\end{multline}
Physically the presence of the two exponentials written in the expansion \eqref{GK_decomposition} represents a reflection from the tip of the cigar. Note that these two exponentials have the same conformal dimension, see \eqref{dimV}.
For our case of type IIA string we will take $m_L = -m_R = m $. 

The reflection coefficient \eqref{refl_GK} is defined up to an arbitrary factor of the form $(A)^{2j + 1}$,  where $A$ is a constant.  It can be absorbed into a redefinition of the operators 
\begin{equation}
	V_{j, m_L, m_R} \mapsto (A)^{j} V_{j, m_L, m_R},
\label{V_redef}
\end{equation}
see  \cite{LSZinLST}. We use the normalization of \cite{Teschner:1997ft,Teschner:1999ug,LSZinLST}.

The primaries under consideration obey the so-called reflection property%
 \cite{LSZinLST}:
\begin{equation}
	V_{j ; m_L, m_R}=  R(j, m_L, m_R ; k) V_{-j-1 ; m_L, m_R}
\label{GK_reflection_property}
\end{equation}
Note a useful relation
\begin{equation}
	R(j, m_L, m_R ; k) \cdot R(-j-1, m_L, m_R ; k) = 1 \,.
\label{refl_inverse}
\end{equation}
The reflection property \eqref{GK_reflection_property} can be easily checked on the level of expansion \eqref{GK_decomposition}.
We have,
\begin{equation}
\begin{aligned}
	V_{j, m_L, m_R}
		&\approx  e^{i Q(m_L Y_L + m_R Y_R)}\left[e^{Q j \phi}+ R(j, m_L, m_R ; k) e^{-Q(j+1) \phi} \right] \\
		&= R(j, m_L, m_R ; k) \  e^{i Q(m_L Y_L + m_R Y_R)}\left[ R(-j-1, m_L, m_R ; k) e^{Q j \phi}+  e^{-Q(j+1) \phi} \right] \\
		&\approx  R(j, m_L, m_R ; k) \ V_{-j-1 ; m_L, m_R},
\end{aligned}
\label{reflection_check}
\end{equation}
where  we  use \eqref{refl_inverse}.

For generic values of $j$ one exponential in 
 \eqref{GK_decomposition} is normalizable while the other one is  non-normalizable, see \eqref{normalizable}, which stays intact for generic values of $k$.
However, at special values of $j, m_L, m_R$ these operators can become normalizable.
This can happen in one of the following cases:
\begin{enumerate}
\item If $j < -1/2$ and the reflection coefficient $R(j, m_L, m_R ; k)$ vanishes;
\item If $j > -1/2$ and the reflection coefficient $R(j, m_L, m_R ; k)$ has a pole;
\item If $j = -1/2$.
\end{enumerate}
To see this, consider an operator $V_{j ; m_L, m_R}$ with $j < -1/2$.
When  $j$ and $m_L, m_R$ approach the values of  the discrete series \eqref{discrete},  the reflection coefficient vanishes.
Technically it happens because some gamma functions in \eqref{refl_GK} develop poles; for more details see below.
Using the expansion  \eqref{GK_decomposition} we find
\begin{multline}
	V_{j ; m_L, m_R} \sim e^{i Q(m_L Y_L + m_R Y_R)} \left( e^{Q j \phi} + \underbrace{R(j, m_L, m_R ; k)}_{\to 0} e^{-Q (j+1) \phi} \right) \to \\
	\to
	e^{Q j \phi + i Q(m_L Y_L + m_R Y_R)},
\label{V_normalizable}
\end{multline}
which is the  leading term in the expansion of a normalizable operator.

For an operator $V_{\tilde{j}}$ with $\tilde{j}=-1-j > -1/2$ we have similarly
\begin{multline}
	V_{\tilde{j} ; m_L, m_R} \sim e^{i Q(m_L Y_L + m_R Y_R)} \left( e^{Q \tilde{j} \phi} + \underbrace{R(\tilde{j}, m_L, m_R ; k)}_{\text{has a pole}} e^{Q j \phi} \right) \to \\
	\to
	R(\tilde{j}, m_L, m_R ; k) e^{Q j \phi + i Q(m_L Y_L + m_R Y_R)}
\label{V_non-normalizable}
\end{multline}
(cf. \eqref{refl_inverse}). 
This expression contains a pole. 
The residue at this pole is again a normalizable operator (see below for more details).

As for $j = -\frac12$, it is a borderline case when the two exponentials in \eqref{GK_decomposition} are the same. 
As we already mentioned it is a state with the norm logarithmically divergent in the infrared.
As was argued in \cite{SYlittles}, this state should be included into the the physical spectrum.

Thus we can have normalizable as well as non-normalizable operators. 

In the following we will be interested in correlation functions of operators corresponding to hadronic states of 4D SQCD.
Strictly speaking, these are given by \textquote{dressed} vertex operators \eqref{tachyon}, \eqref{graviton}, and apart from the ${\rm SL}(2,R)/{\rm U}(1)$ part, they include ghost and ${\mathbb R}^4$ factors. However these \textquote{extra} factors contribute to the correlation functions trivially. 
Below we will focus on the \textquote{internal} $(\phi, Y)$-dependent ${\rm SL}(2,R)/{\rm U}(1)$ part in the correlation functions.
The latter will contain all the relevant pole structure, and only this $(\phi, Y)$-dependent part is relevant for studying normalizable and non-normalizable operators in the sense discussed above.

\subsection{Poles of the two point function and holography}

AdS/CFT-type holography discussed in Sec.~\ref{sec:hologr_views} relates correlation functions of non-normalizable operators with $j > -1/2$  to correlation functions of normalizable operators with $j < -1/2$, see \eqref{LSZ}
and \eqref{V_non-normalizable}.
The reflection property \eqref{GK_reflection_property} and the expansion \eqref{GK_decomposition} are main technical ingredients necessary for understanding this relation.

The two point correlation function of the primary operators was computed in \cite{GivKut} (see also \cite{Teschner:1997ft,Teschner:1999ug} and  \cite{LSZinLST,Giveon:1999tq}). The exact formula reads%
\footnote{%
In this paper we mostly do not write the dependence on the world sheet coordinates explicitly.
Moreover, note the $\delta(j_1 - j_2)$ factor is absent in \cite{LSZinLST,Giveon:1999tq}. The authors of those paper argued that this delta function cancels against the ghost zero mode that would normally make the two point  amplitude vanish in string theory. However according to the recent work \cite{Erbin:2019uiz}, the two point string amplitude are non-vanishing after all. Therefore, we should keep the delta function $\delta(j_1 - j_2)$.%
}
\begin{equation}
	\left\langle V_{j_1 ; m_L, m_R } V_{j_2 ; -m_L, -m_R}\right\rangle
	=
	R(j, m_L, m_R ; k) \, \delta(j_1 - j_2)
\label{2pt_via_refl}
\end{equation}
with the quantity $R$ (the reflection coefficient) given by \eqref{refl_GK}.
A quasiclassical  justification of this  formula \eqref{2pt_via_refl} is presented in Appendix~\ref{sec:2pt_check}.
Note that here and below we will suppress the standard dependence of the world sheet coordinates.

First we note that since the two point correlation function \eqref{2pt_via_refl} equals to the reflection coefficient \eqref{refl_GK}, the poles of the correlation function exactly correspond to the values of $j, m_L, m_R$ for which the operators inside the correlation function become normalizable, see Sec.~\ref{sec:refl}.
Second, when we consider correlation functions of normalizable operators with $j \leqslant -1/2$, at special values of $j, m_L, m_R$ they  turn out to be finite, as we will see below.

Non-normalizable  operator with $j > -1/2$ at special values of $j, m$ reduces to a normalizable operator times a reflection coefficient with a pole. This ensures the relation \eqref{LSZ} found in \cite{LSZinLST}.
Therefore, we can expect that the two point correlation function of two such operators has a pole.


As we explain it  Sec.\ref{sec:hologr_views} we can interpret this  in the spirit of AdS/CFT-type holography which assumes that
off mass-shell correlation functions in the 4D field theory are given by string correlation
functions of non-normalizable operators. Poles are associated with propagation of a
physical state.

Let us reiterate.
Correlation function $\langle V_{\tilde{j}, m_L, m_R}, O_1,...,O_n\rangle$ with a non-normalizable operator $V_{\tilde{j}, m_L, m_R}$ with $\tilde{j} > -1/2$ 
corresponds to 
a correlation function $\langle V_{j, m_L, m_R}, O_1,...,O_n\rangle$ with the normalizable operator 
$V_{j, m_L, m_R}$  with $j < -1/2$ (see \eqref{V_non-normalizable}). Here $O_1,...,O_n$ denote other vertex operators in a correlation function.

Specifically, $j$ and $\tilde{j}$ for these two operators are related as
\begin{equation}
	\underbrace{\tilde{j}}_{\text{non-norm.}}
	=
	- 1 - \underbrace{j}_{\text{norm.}}
\label{AdS-CFT_j}
\end{equation}
As we already mentioned these two vertex operators $V_{\tilde{j}, m_L, m_R}$ and $V_{j, m_L, m_R}$ have the same conformal dimension and we  expect that as $(j,m_L,m_R)$ approach values for the discrete spectrum 
the reflection coefficient develops a pole. Then  two vertex operators $V_{\tilde{j}, m_L, m_R}$ and $V_{j, m_L, m_R}$ satisfy the LSZ relation \eqref{LSZ} which can be called \textquote{{\em the} holography relation}. The pole of the propagator of a 4D physical state
in this relation comes from the pole of the reflection coefficient 
with respect to $(j,m_L,m_R)$ via relation \eqref{tachphys} or \eqref{gravphys}.


\subsection{Poles of the two point function and discrete representations of SL($2, \mathbb{R}$)/U(1)}

Now let us look more closely at pole structure of two point correlation functions.

Poles of the correlation functions of operators living on the covering space of SL($2, \mathbb{R}$) were analyzed in \cite{LSZinLST,Giveon:1999tq}. 
Because of the cover, the values of $m = m_{L}$ in the vertex operators \eqref{GK_decomposition} are not quantized in this setting. Moreover, the SL($2, \mathbb{R}$)-spin $j$ was not restricted to take discrete values in \cite{LSZinLST,Giveon:1999tq}.

In our setup 
the field $Y$ in \eqref{GK_decomposition} in the WZNW formualation lives on a circle of radius $R_{sl} = \sqrt{2k}$
\footnote{Note that the radius of a circle in the Liouville formulation is given by \eqref{Y_period}, while the radius of a circle in the mirror WZNW formulation is given by $R_{sl} = \sqrt{2k}$, see \cite{GivKut}. To simplify notation we use the same notation $Y$ for this periodic field.}.
It is a periodic variable.
We do not consider the covering space.
Instead, we require the primaries to be $2 \pi R_{sl}$ periodic with respect to the field $Y$. This immediately imposes a condition
\begin{equation}
	2 m \in \mathbb{Z} \,,
\label{m_halfintegral}
\end{equation}
for arbitrary $k$.
In the following we will always assume that \eqref{m_halfintegral} is true.

We are going to consider a two-point function \eqref{2pt_via_refl} of two non-normalizable operators with $\tilde{j} = -1-j > -1/2$.
We are interested in poles of this two-point function that depend on both $j$ and $m$ (in \cite{LSZinLST} they were called LSZ poles).
It turns out that the values of $j, m$ corresponding to these poles exactly correspond to discrete representations of SL($2, \mathbb{R}$)/U(1) \eqref{discrete}.
Here we are going to show that the physical poles of the two point correlation functions of non-normalizable operators give exactly the $j=-1$ series.
The operators with $j=-1/2$ are on the borderline between normalizable and non-normalizable, and the corresponding two point function turns out to be finite. 

Note that one should be careful when analyzing potentially divergent quantities. From what has been said so far we can deduce the following rule:
when investigating the poles of $j, m, k$-dependent quantities (e.g. correlation functions), we have to take the limits in the following order:
\begin{enumerate}
\item First take $m$ to be a half-integer \eqref{m_halfintegral}.
\item Next, send $j$ to the desired value $j_0$. Holography relation \eqref{LSZ} suggests that it  corresponds  to the   discrete representation of SL($2, \mathbb{R}$)/U(1), e.g. $j\to -1/2$ or $j \to -1$, see \eqref{j=-1/2} and 
\eqref{j=-1}.
\item Lastly, take the limit $k \to 1$.
\end{enumerate}

This ensures that LSZ ''holography relation'' \eqref{LSZ} suggested in \cite{LSZinLST} can  be written in our theory as follows
\beq
\langle V_{\tilde{j}, m_L, m_R}, O_1,...,O_n\rangle \sim \frac1{j-j_0}\,\langle V_{j, m_L, m_R}, O_1,...,O_n\rangle ,
\label{LSZ1}
\eeq
near the pole at $j\to j_{0}$, where we fixed $m$ to be a (half)integer. Here $\tilde{j}$ and $j$ are related via 
\eqref{AdS-CFT_j}, while   $j_0=-1$. Below we are going to check this relation for the two and three point correlation functions in our theory.

We will also check the holography relation \eqref{LSZ} for the $j_0=-1/2$ channel and show that holography does not work in this case.

\subsubsection{$j = -1$}

Let us consider the two point correlation function  \eqref{2pt_via_refl} with two non-normalizable operators $V_{\tilde{j},m_L, m_R}$ 
near $\tilde{j} = -1-j = 0$.
To investigate the pole structure we can use a technique similar to the one outlined in \cite{Giveon:1999tq} (in particular see eq. (3.10) - (3.13) in that paper).
Let us  take $m_L = - m_R \equiv m > 0$.
Using \eqref{2pt_via_refl} and  expanding the $\Gamma$-functions in \eqref{refl_GK} we can see that the two point correlation function has a pole at $\tilde{j} = 0$,
\begin{equation}
	\left\langle V_{\tilde{j} ; m, -m} V_{\tilde{j}' ;-m,m}\right\rangle
	=
	R_m \,\frac1{\tilde{j}} \,\delta(\tilde{j} - \tilde{j}')
\label{2pt_jtild=0}
\end{equation}
with the residue given by (see Appendix~\ref{sec:useful})
\begin{equation}
	R_m = \, \underset{\tilde{j} = 0}{\Res} R(\tilde{j}, m, -m ; k) = \frac{m^2}{2 \pi}
\label{j=-1_res}
\end{equation}
%
This formula  holds also for $m < 0$.
Recalling that $\tilde{j} = -1-j$ 
we see that the singularity actually appears at
\begin{equation}
	j=-1, \quad m=\pm\{1,2, \ldots\}
\label{j=-1_series_rep}
\end{equation}
and the two-point correlation function \eqref{2pt_jtild=0} can be rewritten as
\begin{equation}
	\left\langle V_{\tilde{j} ; m, -m} V_{\tilde{j}' ;-m,m}\right\rangle
	=
	\frac{R_m}{ - 1 - j }\,\delta(\tilde{j} - \tilde{j}')
	\label{j=-1poles}
\end{equation}
which precisely corresponds to a $j=-1$ discrete series of ${\rm SL}(2,R)/{\rm U}(1)$ representations \eqref{j=-1}.
Note that the residue \eqref{j=-1_res}  vanishes  at  $m=0$ which directly corresponds
to the fact that  $m=0$ associated with the would-be massless 
4D graviton is  excluded from the series \eqref{j=-1}.
Poles in \eqref{j=-1poles} correspond to a part of massive physical states \eqref{gravitonmass} found in \cite{SYlittles}.

Thus, we see that AdS/CFT-type holography  works in our theory for 4D physical states with $j=-1$. Namely,
these 4D SQCD  states are seen as  poles in correlation functions  of non-normalizable vertex operators in the string theory in   accordance with \eqref{LSZ} and \eqref{LSZ1}.

Let us note that at $2m \in \mathbb{Z}$ the two point correlation function also has other poles at $\tilde{j} > 0$ (corresponding to $j < -1$).
Similar  poles are called ''bulk poles'' in \cite{LSZinLST} because they come from the region of large $\phi$ rather then from states localized near the tip of the cigar (which are called LSZ poles), see also  discussions in \cite{GivKut,Maldacena:2000hw,Maldacena:2001km}. In our theory these poles are not physical and should be ignored. 
Associated states with $j< -1$ break the lower bound in \eqref{no_ghosts} and have negative norm.

Finally let us make a note on the  two point function of the corresponding normalizable operators. According to \eqref{2pt_via_refl},
\begin{equation}
	\left\langle V_{j=-1 ; m, -m} V_{ j=-1 ;-m,m} \right\rangle
	=
	R(j=-1, m, -m ; k) \, \delta(0)
\end{equation}
The reflection coefficient here vanishes, but this zero cancels against the delta function, and the whole two point function is finite.

\subsubsection{$j = -1/2$ normalizable}
\label{sec:2pt_j=-1/2}

Now consider series of states with $j=-1/2$ , see eq. \eqref{j=-1/2}.
 These states are on the borderline between normalizable
and non-normalizable and saturate the so-called Seiberg bound \cite{Seiberg:1990eb}). As we already mentioned they are logarithmically normalizable with respect to the conifold radial coordinate and were interpreted as  physical states, see \cite{SYlittles} and
Sec.~\ref{sec:vertices}.
Let us have a closer look at the two point function of these states.

Let us set for a moment $j = -1/2 - \epsilon$, $m_L = -m_R \equiv m = 1/2 + \delta$.
We have:
\begin{equation}
\begin{aligned}
	\big\langle &V_{j ; m, -m} V_{j ;-m,m} \big\rangle \\
	&=	R(j, m, -m ; k) \\
	&= \left[ \frac{1}{\pi} \frac{\Gamma\left(1+\frac{1}{k}\right)}{\Gamma\left(1-\frac{1}{k}\right)} \right]^{2j+1}
		\frac{\Gamma\left(1-\frac{2 j+1}{k}\right) \Gamma(j+m+1) \Gamma(j-m+1) \Gamma(-2 j-1)}%
		{\Gamma\left(1+\frac{2 j+1}{k}\right) \Gamma(m-j) \Gamma(-j-m) \Gamma(2 j+1)}
		\\
	&= \left[ \frac{1}{\pi} \frac{\Gamma\left(1+\frac{1}{k}\right)}{\Gamma\left(1-\frac{1}{k}\right)} \right]^{- 2 \epsilon}
		\frac{\Gamma\left(1 + \frac{ 2 \epsilon}{k}\right) \Gamma(1 - \epsilon + \delta) \Gamma(- \epsilon - \delta) \Gamma(2 \epsilon)}%
		{\Gamma\left(1 - \frac{2 \epsilon}{k}\right) \Gamma(1 + \delta + \epsilon) \Gamma(\epsilon - \delta) \Gamma(- 2 \epsilon)}
		\\
	&= \frac{\epsilon - \delta}{\epsilon + \delta} + (\text{non-singular terms}).
\end{aligned}
\end{equation}
This analysis can be repeated for other values of $m, m \in 1/2 + \mathbb{Z}$ with the same result.
One can see that generally speaking this expression is ambiguous. 
However, by our prescription we have to first set $m = 1/2$ (i.e. $\delta = 0$) and then take the limit $j \to -1/2$ (i.e. $\epsilon \to 0$). Finally we take the limit $k\to 1$.
Then the expression for the two point correlation function is well defined, and we have 
\begin{equation}
	\left\langle V_{- 1/2 ; m, -m} V_{- 1/2 ;-m,m}\right\rangle
	\equiv
	\lim\limits_{j \to -1/2}  \left\langle V_{ j ; m, -m} V_{ j ;-m,m}\right\rangle
	= 1 \,.
\label{borderline_2pt}
\end{equation}
The correlation function is finite. This confirms the interpretation of $j=-\frac12$ states as physical states (logarithmically) localized near the tip of the cigar.

Finally let us note that the factor $\delta(j_1 - j_2)$ coming from \eqref{2pt_via_refl} in fact does not enter the final expression for the two point function for $j=-1/2$. For details see Sec.~\ref{sec:cont_rep}.

\subsubsection{$j = -1/2$ non-normalizable}
\label{sec:2pt_j=-1/2_non-norm}

Since the value $j=-1/2$ is invariant under the reflection \eqref{AdS-CFT_j}, one may think that there is no corresponding non-normalizable operator. 
It turns out that this is not entirely true.
We saw that for each operator $V_j$ with $j \neq -1/2$ there is an operator $V_{\tilde{j}}$, $\tilde{j} = - j - 1$ with the same conformal dimension. This is related to the fact that the Schr{\"o}dinger equation in the corresponding quantum mechanical problem has two solutions. 
In the case of the operator $V_{j=-1/2}$ the second solution to the Schr{\"o}dinger equation is the operator \cite{Seiberg:1990eb}
\begin{equation}
	\phi V_{\tilde{j} = -1/2, m, -m} \sim \phi \, e^{- \frac{Q}{2} \phi} e^{i Q(m Y_L - m Y_R)}	\,.
\label{V_j-1-half_nonnorm}
\end{equation}
Therefore, we should take this operator into account.

The operator \eqref{V_j-1-half_nonnorm} is an example of a so-called logarithmic primary field \cite{Gurarie:1993xq} (see also \cite{Nagi:2005cm,Nivesvivat:2020gdj} and references therein).
Such operators were also considered to some extend in \cite{KutSeib,Kutasov:1999xu}. 
It is a primary field with the same conformal dimension as $V_{j = -1/2, m, -m} \sim e^{- Q/2 \phi} e^{i Q(m Y_L - m Y_R)}$. 
Indeed, consider its pairing with the energy-momentum tensor:
%
\begin{equation}
\begin{aligned}
	&<\phi e^{Q j \phi}e^{i Q(m Y_L - m Y_R)}(z_1), T(z_2)> = \frac{1}{Q} \pt_{j} <e^{Q j \phi}e^{i Q(m Y_L - m Y_R)}(z_1),
	T(z_2)> \\
		&= \frac{1}{(z_1-z_2)^{2}} \ \frac{1}{Q} \pt_{j} \left( \Delta_{j,m} \ e^{Q j \phi}e^{i Q(m Y_L - m Y_R)} + \ldots \right) \\
		&= \frac{e^{i Q(m Y_L - m Y_R)}}{(z_1-z_2)^{2}} \left( -\frac{1}{Q} \frac{2j+1}{k}  e^{Q j \phi} 
		+ \Delta_{j,m} \, \phi \, e^{Q j \phi} \right) \Bigg|_{j=-\frac12} +... \\
		&= \frac{\Delta_{j,m}}{(z_1-z_2)^{2}}\,\phi \, e^{Q j \phi}e^{i Q(m Y_L - m Y_R)}  + \ldots,
\end{aligned}
\end{equation}
where $z_1$ and $z_2$ are world sheet coordinates. This shows that
operator 
$$\phi \, V_{\tilde{j} = -1/2, m, -m} \sim \phi e^{- \frac{Q}{2} \phi} e^{i Q(m Y_L - m Y_R)}$$ 
is a primary field with conformal dimension $\Delta_{j=-1/2,m}$, where 
 $\Delta_{j,m}$ is defined in \eqref{dimV}. The associated wave function is more divergent at large $\phi$ than the one for
$V_{j=-1/2}$, therefore we will consider the operator \eqref{V_j-1-half_nonnorm} as a non-normalizable ''partner'' of 
$V_{j=-1/2}$.

The corresponding two point correlation function turns out to have a double pole (see Appendix~\ref{sec:log_2pt} for the details),
\begin{equation}
	\langle \phi V_{\tilde{j}_1 = -1/2, m, -m} \ \phi V_{\tilde{j}_2 = -1/2, -m, m} \rangle
		\sim \frac{1}{(\tilde{j}_1 + \frac{1}{2})^2} \delta(\tilde{j}_1 - \tilde{j}_2)
\label{V_j-1-half_2pt_result}
\end{equation}
The double pole is exactly what one would expect for the correlation function \eqref{V_j-1-half_2pt_result} from LSZ.	
From \eqref{tachphys} we see that  the linear in $(\tilde{j}+1/2)$ term vanishes at $\tilde{j}\to -1/2$ and we have
\begin{equation}
	p_{\mu}^2 + M^2 \sim \left( \tilde{j} + \frac{1}{2} \right)^2 
	\text{  near  } \tilde{j} \to -\frac{1}{2}
\label{p2-M2_jonehalf}
\end{equation}
Therefore  finally we get  the LSZ formula
\begin{equation}
	\langle \phi V_{\tilde{j}_1 = -1/2, m, -m} \ \phi V_{\tilde{j}_2 = -1/2, -m, m} \rangle
		\sim \frac{1}{p_{\mu}^2 + M^2} \langle  V_{j = -1/2, m, -m} \  V_{j = -1/2, -m, m} \rangle .
\end{equation}

We conclude that the LSZ pole is present   for the non-normalizable  $j=-1/2$ operator 
\eqref{V_j-1-half_nonnorm}, so the holography works in this channel for the two-point function.
Below we will see that it does not work for the three point function of such operators.

%
%

\section{Three point correlation functions}
\label{sec:3pt_VVV}

In this section we consider three point correlation functions of  operators $V_{j,m_L, m_R}$ with $m_L = - m_R \equiv m$.
The physical states at our disposal correspond to normalizable operators with $j=-1, -1/2$, or by \eqref{AdS-CFT_j} to $\tilde{j} = 0, -1/2$.
The values of $m$ of the three correlation functions have to sum up to zero,
\begin{equation}
	m_1 + m_2 + m_3 = 0
\label{3pt_winding_conservation}
\end{equation}
This corresponds to baryon charge conservation in the 4D SQCD, see \eqref{m-baryon}.
Winding number conservation in three point functions was also discussed in \cite{Maldacena:2001km}. 

Therefore we have two potentially non-vanishing three point functions: 
$j = (-1,-1/2,-1/2)$ and $j = (-1,-1,-1)$. 
In the following we are going to investigate these two cases.

Let us start with  general comments. The holography relation \eqref{LSZ1} suggests that
 the correlation functions of non-normalizable operators should be singular, 
with the poles corresponding to propagators of  normalizable states (discrete ${\rm SL}(2,R)/{\rm U}(1)$ representations),
The residues are related to the correlation functions of normalizable operators. The latter correlation functions are finite.

The three point correlation functions for the coset SL($2, \mathbb{R}$)/U(1) model were computed in \cite{Giveon:1999tq}, see also \cite{Teschner:1997ft,Teschner:2001rv}.
The pole structure can be investigated by the method used in \cite{Giveon:1999tq}.
(Note however that there is a typo in \cite[eq. (4.20)]{Giveon:1999tq}), see Appendix~\ref{sec:useful}).

\subsection{$j = (-1,-1/2,-1/2)$ correlation function}
\label{sec:3pt_VVV_decay}

In this subsection we are going to consider the three point correlation function with one $j=-1$ operator and two $j=-1/2$ operators, and also investigate its possible holographic relation to the correlation function with corresponding non-normalizable operators.
We will see that there is indeed such a relation in the $j=-1 \leftrightarrow \tilde{j}=0$ channel. Let us start with this one.

\subsubsection{A non-normalizable $\tilde{j}_1 = 0$ operator}

Consider three operators with SL($2, \mathbb{R}$) spins $\tilde{j}_1 = 0,\ j_2 = -1/2,\ j_3 = -1/2$.
The first one is non-normalizable and by \eqref{AdS-CFT_j} it should correspond to a normalizable operator with $j_1 = -1$).
Since the other two operators are not non-normalizable, they do not lead to poles in the three point function, and the resulting correlation function contains only one pole.
By using the technique of \cite{Giveon:1999tq} (see Appendix~\ref{sec:3pt_discr_calculation} for details) we  arrive at the expression
\begin{equation}
	\left\langle  V_{\tilde{j}_{1} ; m_{1}, -m_{1}} V_{j_{2}=-1/2 ; m_{2}, -m_{2}} V_{j_{3}=-1/2 ; m_{3}, -m_{3}} \right\rangle
	= \frac{R_{m_1 m_2 m_3}}{ \tilde{j}_{1} }
		+ \text{(regular terms)}
\label{3pt_V-correlator}
\end{equation}
where we used that $m_1 \in \mathbb{Z} \backslash \{ 0 \}$ and $m_2, m_3 \in 1/2 + \mathbb{Z}$. We see that there is a pole at $\tilde{j}_{1} = 0$.
The residue is calculated to be
\begin{equation}
	R_{m_1 m_2 m_3} = \frac{1}{2\pi}
\label{3pt_V-correlator_res}
\end{equation}
cf. \eqref{3pt_V-correlator_res_with_normalizable} and the comments below.

\subsubsection{A normalizable $j=-1$ operator}
\label{sec:3pt_VVV_decay_normalizable}

Now we consider a similar correlation function, only this time with all fields normalizable,  $j_1 = -1, j_2 = j_3 = -1/2$.
Namely, consider the three-point correlation function
\begin{equation}
	\left\langle  V_{j_1=-1 ; m_{1}, -m_{1}} V_{j_{2}=-1/2 ; m_{2}, -m_{2}} V_{j_{3}=-1/2 ; m_{3}, -m_{3}} \right\rangle .
\label{3pt_1-half-half_normalizable}
\end{equation}
Using the reflection property \eqref{GK_reflection_property} we obtain the exact expression
\begin{equation}
\left\langle  V_{j_1=-1 ; m_{1}, -m_{1}} V_{j_{2} ; m_{2}, -m_{2}} V_{j_{3} ; m_{3}, -m_{3}} \right\rangle
	= 
	\frac{ \left\langle  V_{\tilde{j}_1=0 ; m_{1}, -m_{1}} V_{j_{2} ; m_{2}, -m_{2}} V_{j_{3} ; m_{3}, -m_{3}} 
	\right\rangle }{R(\tilde{j}_1=0, m_1, -m_1 ; k)}
\end{equation}
As we saw above, both the numerator and denominator have poles at $\tilde{j}_1 = - 1 - j_1 = 0$. Substituting 
\eqref{3pt_V-correlator_res} and \eqref{j=-1_res} here we get
\begin{equation}
\begin{aligned}
\big\langle  V_{j_1=-1 ; m_{1}, -m_{1}} &V_{j_{2}=-1/2 ; m_{2}, -m_{2}} V_{j_{3}=-1/2 ; m_{3}, -m_{3}} \big\rangle \\
	&= 
	\frac{ \underset{\tilde{j}_1=0}{\Res} \left\langle  V_{\tilde{j}_1 ; m_{1}, -m_{1}} V_{j_{2}=-1/2 ; m_{2}, -m_{2}} V_{j_{3}=-1/2 ; m_{3}, -m_{3}} \right\rangle }
	{ \underset{\tilde{j}_1=0}{\Res} R(\tilde{j}_1, m_1, -m_1 ; k)}
	\\
	&= \frac{1}{m_1^2}
\end{aligned}
\label{3pt_V-correlator_res_with_normalizable}
\end{equation}
This correlation function is finite. Note, that the value $m_1 = 0$ is excluded by \eqref{j=-1}.

To conclude this subsection, let us note that the correlation function \eqref{3pt_1-half-half_normalizable} describes a decay of one 4D physical state into two other physical states. Clearly, the conservation of 4D momentum and mass-shell conditions for all three states ensure that the heaviest state can decay into two other states only if its mass is larger that the sum of masses of 
other states. For example, the state with $j=-1$ can decay into two states from $j=-1/2$ series if its mass is larger then 
the sum of the masses of the two $j=-1/2$ states,
\begin{equation}
	\text{Mass}_{j_1=-1, m_1} \geqslant \text{Mass}_{j_2=-1/2, m_2} + \text{Mass}_{j_3=-1/2, m_3}
\end{equation}
Using mass formulas \eqref{tachyonmass} and \eqref{gravitonmass} and the winding conservation \eqref{3pt_winding_conservation}, we arrive at
\begin{equation}
\begin{aligned}
	m_1^2 &\geqslant m_2^2 - \frac{1}{4} + m_3^3 - \frac{1}{4} + 2 \sqrt{(m_2^2 - \frac{1}{4}) (m_3^3 - \frac{1}{4})} \\
	\Leftrightarrow m_2 \, m_3 &\geqslant - \frac{1}{4}
\end{aligned}
\end{equation}
In our case this is equivalent to the condition that $m_2$ and $m_3$ are of the same sign. If this condition is not satisfied, we have the heaviest $j=-1/2$ state decaying to a lighter $j=-1/2$ state and a $j=-1$ state.
Fig.~\ref{fig:decay_threshold_1} summarizes these results.

\subsubsection{$\phi e^{- Q/2 \phi}$ and would-be LSZ poles}

In section \ref{sec:2pt_j=-1/2_non-norm} we considered the non-normalizable operator with $j=-1/2$. 
We saw that the two point function for such an operator has a double pole.

Turns out that the three point function with insertions of the logarithmic primaries \eqref{V_j-1-half_nonnorm} has an unexpected structure.
This three point function is calculated in Appendix~\ref{sec:3pt_discr_calculation}. The result is
\begin{equation}
\begin{aligned}
	&\left\langle V_{\tilde{j}_1 = 0, m, -m} \ \phi V_{\tilde{j}_2 = -1/2, m, -m} \ \phi V_{\tilde{j}_3 = -1/2, m, -m} \right\rangle \\
	&= \frac{1}{2\pi} \frac{1}{ \tilde{j}_1 \, (\tilde{j}_2 + \frac{1}{2}) \, (\tilde{j}_3 + \frac{1}{2}) }
\end{aligned}
\label{3pt_log}
\end{equation}
This is not the expected LSZ behavior. From \eqref{LSZ}, \eqref{p2-M2_jonehalf} one could expect that the three point correlation function \eqref{3pt_log} has a single pole corresponding to $\tilde{j}_1 = 0$ and two double poles corresponding to $\tilde{j}_{2}, \tilde{j}_3 = -1/2$.
The single pole at $\tilde{j}_1 = 0$ shows up in \eqref{3pt_log} as expected, but the other two are single rather than double poles.
What's going on? 

Moreover,   single  poles at $\tilde{j}_{2}, \tilde{j}_3 = -1/2$ in \eqref{3pt_log} are not LSZ poles. 
Indeed, according to \cite{LSZinLST}, when we consider the $x_i$-dependent correlation function and integrate over $x_i$ (cf. \eqref{3pt_GK_exact_Ffactor}), the LSZ poles are seen as poles coming from $x_i \to 0, \infty$. However this integrated correlation function also exhibits poles that can be traced back to coinciding $x_i$ under the integral (and also to $x \to 1$ in the expression \eqref{3pt_GK_exact_Ffactor}). 
The analysis shows that the single $j=-1/2$ poles in \eqref{3pt_log} are exactly of this type, see Appendix~\ref{sec:3pt_cont_calculation} for the details. 

In other words these are ''bulk poles'' coming from the bulk of the cigar, i.e. from the region of large $\phi$
rather than LSZ poles, which come from states localized near the tip of the cigar.

We  conclude that the holography picture does not work for the $j=-1/2$ channel.

\subsection{$j = (-1,-1,-1)$ correlation function}

Consider three operators with spins $j_1 = j_2 = j_3 = -1$. 
Corresponding non-normalizable operators should have $\tilde{j}_1 = \tilde{j}_2 = \tilde{j}_3 = 0$.

In this case one could expect that the correlation function of non-normalizable operators with $\tilde{j} = 0$ should contain three poles, but it does not.
This can be seen by analyzing the exact formula for the three point function \eqref{3pt_GK_exact}.
The coefficient $D$ from \eqref{3pt_GK_exact_Dfactor} becomes
\begin{equation}
\begin{aligned}
	D &= \frac{k}{2 \pi^3} \left[ \frac{1}{\pi} \frac{\Gamma(1 + \frac{1}{k})}{\Gamma(1 - \frac{1}{k})} \right]^{j_1 + j_2 + j_3 +1}
			\frac{\Gamma \left( 1 - \frac{j_1 + j_2 + j_3 +1}{k} \right)}{\Gamma \left(\frac{j_1 + j_2 + j_3 +1}{k} \right)}
			\\
		&\xrightarrow{ j_i \to 0,\ k\to 1} \frac{k}{2 \pi^4}
\end{aligned}
\end{equation}
The factor \eqref{3pt_GK_exact_Ffactor} cannot produce all the necessary poles, and the correlation function under consideration does not have the necessary pole structure.
See also a discussion in Appendix B of \cite{Giveon:1999tq}.

Similarly we can check that the three point function of three normalizable operators with $j = -1$ is zero.
Although this time the coefficient $D$ in \eqref{3pt_GK_exact_Dfactor}  happens to contain a single pole, 
the function $F(j_1, m_1; j_2, m_2; j_3, m_3 )$  in \eqref{3pt_GK_exact_Ffactor} has a double zero unless $m_1 = m_2 = 0$ \cite{Giveon:1999tq}; however the latter values of $m$ are excluded, see \eqref{j=-1}.
Therefore the whole correlation function vanishes:
\begin{equation}
	\left\langle  V_{j_1=-1 ; m_{1}, -m_{1}} V_{j_{2}=-1 ; m_{2}, -m_{2}} V_{j_{3}=-1 ; m_{3}, -m_{3}} \right\rangle
	= 0.
\end{equation}

This completes the analysis of the three point correlation functions of vertex operators corresponding to baryons of the 4D \ntwo SQCD.

%
%

\section{Continuous representations}
\label{sec:cont_rep}

So far we have discussed only the vertex operators from the discrete representations \eqref{discrete}, as they have direct interpretation as hadrons of the underlying 4D \ntwo SQCD \cite{SYlittles,SYlittmult}. However, there is also the principal and exceptional continuous representations. 

In this section we concentrate on the vertex operators from the principal continuous representation with%
\footnote{On the interpretation of $\Im{j}$ as a momentum along the cigar see e.g.  \cite{Ginsparg:1993is,DiFrancesco:1997nk,Maldacena:2000kv,SYlittles}.}
 $j = -1/2 + is$, see \eqref{principal}. 
The scaling dimensions of these vertex operators are given by
\begin{equation}
	\Delta = m^2 - \frac{1}{4} + s^2
\label{principal_mass}
\end{equation}
The parameter $s \in {\mathbb R}$ here is continuous, which leads to continuous mass spectrum for the states created by these vertex operators.
This prevents interpretation of such states as baryons in 4D  SQCD. It was argued in \cite{SYlittles} that these string states are multiparticle states
 related to the presence of massless bi-fundamental quarks in 4D SQCD. 
To shed more light on their nature below we 
 study correlation functions with insertions of such operators.

\subsection{Two point correlation function}

The two-point correlation function of such states is given by
\begin{equation}
\begin{aligned}
	&\left\langle  V_{ j_1 = -1/2 + is ; m_L, m_R } V_{ j_2 = - 1/2 + is ;-m_L, -m_R}\right\rangle
	=	R(- 1/2 + is, m_L, m_R ; k) \\
	&= 	\left[ \frac{1}{\pi} \frac{\Gamma\left(1+\frac{1}{k}\right)}{\Gamma\left(1-\frac{1}{k}\right)} \right]^{2is}
			\frac{\Gamma\left(1-\frac{2is}{k}\right) \Gamma(1/2 + is +m_L) \Gamma( 1/2 + is + m_R) \Gamma(- 2is)}%
			{\Gamma\left(1+\frac{2is}{k}\right) \Gamma(m_L + 1/2 - is ) \Gamma(1/2 - is + m_R) \Gamma(2is)} \\
	&= \left[ \frac{1}{\pi} \frac{\Gamma\left(1+\frac{1}{k}\right)}{\Gamma\left(1-\frac{1}{k}\right)} \right]^{2is} 
			\cdot e^{i \delta(s; m_L, m_R ; k)} \,,
\end{aligned}
\label{2pt_via_refl-continuous}
\end{equation}
where we have introduced a phase $\delta(s; m_L, m_R ; k)$.
This correlation function does not have poles in the variable $s$; instead, it should be interpreted simply as a scattering phase, 
see e.g. \cite{Maldacena:2001km}.
The only possible problem with \eqref{2pt_via_refl-continuous} is that now the limit $k \to 1$ is not well-defined because of the first factor. 
However this factor is non-physical and can be absorbed into the definition of vertex operators \eqref{V_redef}. 

Finally let us comment on the factor $\delta(j_1 - j_2)$ coming from \eqref{2pt_via_refl}.
It does not enter the formula for the two point function \eqref{2pt_via_refl-continuous} for the following reason.
The states with $j = -1/2 + is$ form a continuum. When we calculate an amplitude with such states, we should integrate over all final states, that is, 
\begin{multline}
	\int ds \, \delta(s - s') 	\left\langle  V_{ j_1 = -1/2 + is ; m_L, m_R } V_{ j_2 = - 1/2 + is' ;-m_L, -m_R}\right\rangle
	\\
	=
	\left\langle  V_{ j_1 = -1/2 + is ; m_L, m_R } V_{ j_2 = - 1/2 + is ;-m_L, -m_R}\right\rangle
\end{multline}
The two point function of $j=-1/2$ operators considered in Sec.~\ref{sec:2pt_j=-1/2} can be considered as the limiting $s \to 0$ case of the correllation function considered here. Therefore, the delta function factor is absent in that expression, too.

\subsection{Three point correlation function}
\label{sec:3pt_cont}

In Sec.~\eqref{sec:3pt_VVV_decay} we derived the three point function with $j=(-1,-1/2,-1/2)$.
That derivation can be easily adopted for the three point function with continuous representation,
which turns out to be non-trivial in the case\footnotemark
\begin{equation}
	j_1 = -1 \leftrightarrow \tilde{j}_1 = 0 \,;
	\quad
	j_2 = - \frac{1}{2} + is \,,
	\quad
	j_3 = - \frac{1}{2} - is \,.
\label{j_vals_3pt_cont}
\end{equation}
\footnotetext{%
%
The case with $j_3 = - \frac{1}{2} + is$ is also non-trivial but can be recovered from \eqref{j_vals_3pt_cont} using the reflection property \eqref{GK_reflection_property}.
%
.%
}
The three point function in this case is as follows (see Appendix~\ref{sec:3pt_cont_calculation} for a detailed computation):
\begin{equation}
	\left\langle  V_{\tilde{j}_{1} \to 0 ; m_{1}, - m_{1}} V_{-1/2 + is ; m_{2}, - m_{2}} V_{-1/2 - is ; m_{3}, - m_{3}} \right\rangle
	=
	\frac{1}{\tilde{j}_1}
	\times 	\frac{1}{2 \pi} e^{ i (\delta(s; m_2, -m_2 ; k) - \delta(s; m_3, -m_3 ; k)) }
\label{3pt_V-correlator_cont_j=0}	
\end{equation}
The phase factor $\delta$ is defined in \eqref{2pt_via_refl-continuous}.
In the limit $s \to 0$ in the three point function \eqref{3pt_V-correlator_cont_j=0} the phase goes to zero and 
we recover the answer \eqref{3pt_V-correlator} with the residue given by \eqref{3pt_V-correlator_res}.

We see that the three point correlation function with one non-normalizable operator with $\tilde{j}=0$ again has a single pole. This confirms the LSZ holography relation \eqref{LSZ1} for correlation functions of operators from principal continues representation.
The corresponding three point function with normalizable operators is going to be finite; it can be calculated by the same method as in 
Sec.~\ref{sec:3pt_VVV_decay_normalizable}. The answer is
\begin{equation}
	\left\langle  V_{-1 ; m_{1}, - m_{1}} V_{-1/2+is ; m_{2}, - m_{2}} V_{-1/2-is ; m_{3}, - m_{3}} \right\rangle
	= \frac{1}{m_1^2} e^{ i (\delta(s; m_2, - m_2 ; k) - \delta(s; m_3, - m_3 ; k)) }
\label{3pt_V-correlator_cont_j=-1}
\end{equation}

\begin{figure}[h]
	\centering
	\includegraphics[width=0.6\linewidth]{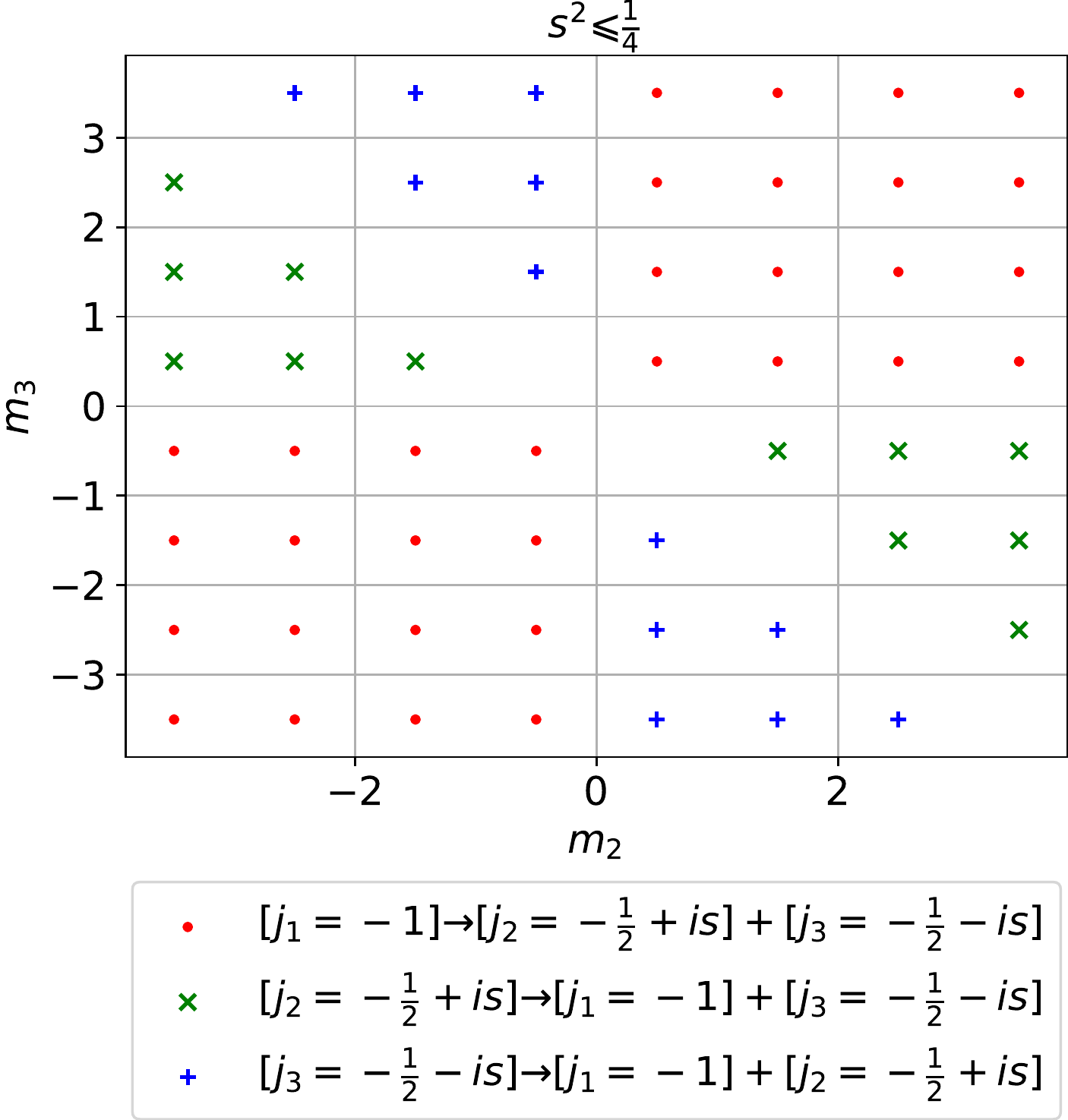}
	\caption{
		Possible decays determined by the energy conservation.
		The quantum numbers $j, m$ are as they appear in the three point correlation function\eqref{3pt_V-correlator_cont_j=-1}.
		The $(m_2, m_3)$ plane extends infinitely.
		Note that the points with $m_2 = - m_3$ are excluded because of winding conservation \eqref{3pt_winding_conservation} and the fact that for $j_1=-1$ it is required that $m_1 \neq 0$ \eqref{gravitonmass}.
	}
\label{fig:decay_threshold_1}
\end{figure}


Much in the same way as for states from the discrete spectrum the decay of $j=-1$ state into two states from $j=-\frac12 \pm is$
series can go only if the mass of the decaying particle is greater than the sum of the masses of the products,
\begin{equation}
	\text{Mass}_{j_1=-1, m_1} \geqslant \text{Mass}_{j_2=-1/2+is, m_2} + \text{Mass}_{j_3=-1/2-is, m_3}
\end{equation}
Using mass formulas \eqref{principal_mass} and \eqref{gravitonmass} and the winding conservation \eqref{3pt_winding_conservation}, we arrive at
\begin{equation}
\begin{aligned}
	m_1^2 &\geqslant m_2^2 + s^2 - \frac{1}{4} + m_3^3 + s^2 - \frac{1}{4} + 2 \sqrt{(m_2^2 + s^2 - \frac{1}{4}) (m_3^3 + s^2 - \frac{1}{4})} \\
	&\Leftrightarrow 
	m_2 m_3 - s^2 + \frac{1}{4} \geqslant \sqrt{(m_2^2 + s^2 - \frac{1}{4}) (m_3^3 + s^2 - \frac{1}{4})} \\
	&\Leftrightarrow 
	\begin{cases}
		s^2 \leqslant \frac{1}{4}  \\
		m_2, m_3 \text{ are of same sign}
	\end{cases}
\end{aligned}
\end{equation}
The condition that the $j_1=-1$ state decays into two continuous with $j_{2,3} = -1/2 \pm is$ turns our to be
\begin{equation}
	\begin{cases}
		s^2 \leqslant \frac{1}{4}  \\
		m_2, m_3 \text{ are of same sign} 
	\end{cases}
\label{j2_mass_threshold}
\end{equation}
Fig.~\ref{fig:decay_threshold_1} summarizes these results.
Note that decay via a three-particle process (when one particles decays to two) is possible if and only if when it involves either all discrete states, or two continuous with $s^2 \leqslant 1/4$. 
A decay via three particle process cannot involve continuous states with $s^2 > 1/4$.

\subsection{Interpretation of continuous representations}
\label{sec:decay}

The correlation function \eqref{3pt_V-correlator_cont_j=-1} that we just computed corresponds to a process where a $j = -1$ state decays into two conjugate states:
\begin{equation}
	[j=-1] \to [j=-\frac{1}{2} + is] + [j=-\frac{1}{2} - is]
\label{reaction_j=1_j=is_j=is}
\end{equation}
cf. the discussion of operator mixing in \cite{Maldacena:2001km}.

How can we interpret the states corresponding to the principal continuous representation \eqref{principal}?
As we already mentioned according to \eqref{principal_mass} their energies form continuous spectrum.
However, in 4D SQCD we do not expect to have a family of hadrons with continuous distribution of masses.

In \cite{SYlittles} it was suggested that these states can be interpreted as decaying modes of normalizable physical 4D states.
This was confirmed by showing that spectra of continues states start from thresholds given by masses \eqref{tachyonmass}
and \eqref{gravitonmass}.
Here we can confirm this interpretation for the case of principal continuous representation. 

Moreover, we can further specify this: a $j= -1/2 + is$ state can be interpreted as a multipartical   state of a $j=-1/2$ baryon and massless bi-fundamental quarks.
Then the decay \eqref{reaction_j=1_j=is_j=is}  corresponds to the decay of $j=-1$ state into two $j=-1/2$ states (which we know can happen from Sec.~\ref{sec:3pt_VVV_decay}) with the  radiation of massless bi-fundamental quarks.

%
%

\section{Conclusions} 
\label{sec:conclusions}

In this paper we  studied correlation functions in the string theory for the critical non-Abelian vortex.
Specifically, we considered  these correlation function using the equivalent description in terms of string theory on 
${\rm SL}(2,\mathbb{R})/{\rm U}(1)$ WZNW coset and studied  their analytic structure.

We also compared our solitonic string-gauge duality, which relates hadrons of 4D \ntwo SQCD to 
closed string states of the string theory of the critical non-Abelian vortex with the AdS/CFT approach. Suggesting that the 
solitonic string-gauge duality can be thought of as  a ''no branes'' limit of the AdS/CFT we tested the holography, which is a distinctive feature of the AdS/CFT correspondence. 

We showed that the AdS/CFT-type holography relation \eqref{LSZ} found for LST \cite{LSZinLST} or \eqref{LSZ1} is fulfilled in our theory for most channels.
This relation ensures that correlation functions of non-normalizable operators in the string theory on the cigar correspond to correlation functions of normalizable operators, which in turn are directly associated with hadronic states in 4D \ntwo SQCD.
The only exception are operators of $j=-1/2$ series, which are on the borderline between normalizable and non-normalizable operators and correspond to physical states in 4D SQCD. It turns out that in the $j=-1/2$ channel holography relation does not work.

Technically the reason for the holography relation \eqref{LSZ} is that non-normalizable and normalizable operators with the same conformal dimension are related by the reflection of the tip of the cigar. It would be interesting to understand a deep
conceptual reason behind this holography.

We can also mention several other open questions. 
The first is to find a more direct relation between critical string theory on the conifold and non-critical $c=1$ string theory with the Liouville field.
The second is the construction of the effective theory of hadron interactions in the 4D \ntwo SQCD. In principle, it should be possible using  results obtained in this paper for correlation functions in the string theory. They fix the coupling constants of the effective interactions of hadrons in SQCD. The symmetry restrictions, especially the \ntwo supersymmetry, heavily restrict the form of possible interaction terms.
Moreover, the interpretation of continuous representations deserves more solid justification.
We leave these problems for a future work.

%
%

\section*{Acknowledgments}

The authors are grateful to M. Shifman for valuable discussions.
%
%
The work of A.Y. was  supported by William I. Fine Theoretical Physics Institute,   
University of Minnesota.

%
%

\clearpage
\appendix

\begin{appendices}

%
%

\section{Three point functions}
\label{sec:3pt_additional}

Now let us review the results for the three point functions in the supersymmetric Liouville theory.

\subsection{Full three point function formula}

The full analytical expression for the three point function of the operators $V_{j, m, -m}$ was derived in \cite{Giveon:1999tq}.
Let us write it down:
\begin{equation}
\begin{aligned}
	\left\langle  V_{\tilde{j}_{1} ; m_{1}, -m_{1}} V_{j_{2} ; m_{2}, -m_{2}} V_{j_{3} ; m_{3}, -m_{3}} \right\rangle
	=
	D(\tilde{j}_1, j_2, j_3; k) \times \\
	F(\tilde{j}_1, m_1; j_2, m_2; j_3, m_3 )
	\int d^{2} x|x|^{2\left(m_{1}+m_{2}+m_{3}-1\right)}
\end{aligned}
\label{3pt_GK_exact}
\end{equation}
The last factor here simply enforces the baryon charge conservation 
\begin{equation}
	m_1 + m_2 + m_3 = 0	
\label{3pt_m_conserv}
\end{equation}
The first factor%
\footnote{The structure constants \eqref{3pt_GK_exact_Dfactor} are equivalent to \eqref{3pt_Tech_1}. They can be derived from \eqref{3pt_Tech_1} using \eqref{G_vs_Upsilon} and slightly changing the normalization of the operators.}
 equals to
\begin{equation}
\begin{aligned}
	&D(\tilde{j}_1, j_2, j_3; k) = 
	\frac{k}{2 \pi^3} \left[ \frac{1}{\pi} \frac{\Gamma\left(1+\frac{1}{k}\right)}{\Gamma\left(1-\frac{1}{k}\right)} \right]^{\tilde{j}_{1} + j_2 + j_3 +1}
	\times \\
	&\frac{G\left(-\tilde{j}_1-j_{2}-j_{3}-2\right) G\left(j_{3}-\tilde{j}_1-j_{2}-1\right) G\left(j_{2}-\tilde{j}_1-j_{3}-1\right) G\left(\tilde{j}_1-j_{2}-j_{3}-1\right)}{G(-1) G\left(-2 \tilde{j}_1-1\right) G\left(-2 j_{2}-1\right) G\left(-2 j_{3}-1\right)}
\end{aligned}
\label{3pt_GK_exact_Dfactor}
\end{equation}
Some properties of the special $G$-function used here are listed in Appendix~\ref{sec:3pt_susy}.
The $F$-factor in \eqref{3pt_GK_exact} is given by
\begin{equation}
\begin{aligned}
	F(\tilde{j}_1, m_1; j_2, m_2; j_3, m_3 )
	=
	\int d^{2} x_{1} d^{2} x_{2}\left|x_{1}\right|^{2\left(\tilde{j}_1+m_{1}\right)}\left|x_{2}\right|^{2\left(j_{2}+m_{2}\right)} \times
	\\
	\left|1-x_{1}\right|^{2\left(j_{2}-\tilde{j}_1-j_{3}-1\right)}
	\left|1-x_{2}\right|^{2\left(\tilde{j}_1-j_{2}-j_{3}-1\right)}
	\left|x_{1}-x_{2}\right|^{2\left(j_{3}-\tilde{j}_1-j_{2}-1\right)}
\end{aligned}
\label{3pt_GK_exact_Ffactor}
\end{equation}

\subsection{The case with continuous and discrete representations}

Now let us calculate the three point functions \eqref{3pt_V-correlator_cont_j=0} and \eqref{3pt_V-correlator}.

\subsubsection{$j = -1/2 \pm is$}
\label{sec:3pt_cont_calculation}

Now let us turn to the computation of the three point function in the case \eqref{j_vals_3pt_cont},
\begin{equation}
	j_1 = -1 \leftrightarrow \tilde{j}_1 = 0 \,;
	\quad
	j_2 = - \frac{1}{2} + is \,,
	\quad
	j_3 = - \frac{1}{2} - is \,.
\label{j_vals_3pt_cont_appendix}
\end{equation}
We are going to analyze the exact expression for the three point function \eqref{3pt_GK_exact}.
Substituting \eqref{j_vals_3pt_cont_appendix} into \eqref{3pt_GK_exact_Dfactor} and using \eqref{G_properties_GK} and \eqref{G_property_1} we obtain
\begin{equation}
\begin{aligned}
	D\left(\tilde{j}_1, j_2, j_3; k\right)
		&= \frac{k}{2 \pi^3} \times \frac{G(-1) G(-2is-1) G(2is-1) G(\tilde{j}_1)}{G(-1) G(-1) G(-2is) G(2is)} + \ldots \\
		&= \frac{k}{2 \pi^3} \left( - \frac{k}{\tilde{j}_1} \right) \left( - \frac{(2is)^2}{ k^2 } \right) + \ldots \\
		&= - \frac{2 s^2 }{\pi^3} \, \frac{1}{\tilde{j}_1} + \ldots
\end{aligned}
\end{equation}
where the ellipses stand for the terms non-singular at \eqref{j_vals_3pt_cont}.
Next, let us evaluate the integral factor \eqref{3pt_GK_exact_Ffactor}. 
From \eqref{j_vals_3pt_cont} we see that $\tilde{j}_1-j_{2}-j_{3}-1 = 0$. 
Making a change of variables 
\begin{equation}
	x_1 = x, \quad  x_2 = x \, t
\end{equation}
we can rewrite $F(\tilde{j}_1, m_1; j_2, m_2; j_3, m_3 )$ from \eqref{3pt_GK_exact_Ffactor} as
\begin{equation}
\begin{aligned}
	F(\tilde{j}_1, m_1; j_2, m_2; j_3, m_3 )
	&= \int d^2 x d^2 t
	\\
	&|x|^{2 (\tilde{j}_1+m_1) + 2 + 2 (j_2 + m_2) + 2 (j_{3}-\tilde{j}_1-j_{2}-1)}  
	|1 - x|^{2 (j_2 - \tilde{j}_1 - j_3 - 1)}
	\\
	&\times |t|^{2 (j_2 + m_2)}
	|1 - t|^{2 (j_{3}-\tilde{j}_1-j_{2}-1)}
\end{aligned}
\end{equation}
Using \eqref{3pt_m_conserv}, we obtain
\begin{equation}
\begin{aligned}
	F(\tilde{j}_1, m_1; j_2, m_2; j_3, m_3 )
	&= \int d^2 x d^2 t
	\\
	&|x|^{2 (j_3 - m_3)}
	|1 - x|^{2 (j_2 - \tilde{j}_1 - j_3 - 1)}
	\\
	&\times |t|^{2 (j_2 + m_2)}
	|1 - t|^{2 (j_{3}-\tilde{j}_1-j_{2}-1)}
\end{aligned}
\end{equation}
Finally, substituting \eqref{j_vals_3pt_cont} here gives
\begin{equation}
\begin{aligned}
	F(\tilde{j}_1, m_1; j_2, m_2; j_3, m_3 )
	&= \int d^2 x 
	|x|^{2 (-1/2 - m_3 - is)}
	|1 - x|^{2 (2is - 1)}
	\\
	&\times \int d^2 t
	|t|^{2 (-1/2 + m_2 + is)}
	|1 - t|^{2 (- 2is -1)}
\end{aligned}
\end{equation}
These integrals can be calculated with the help of \eqref{master_integral}:
\begin{equation}
\begin{aligned}
	\int d^2 x & |x|^{2 (-1/2 - m_3 - is)} |1 - x|^{2 (2is - 1)} \\
		&= \pi \frac{\Gamma(1/2 - m_3 - is) \Gamma(2is) \Gamma(-is + m_3 + 1/2)}
					{\Gamma(1/2 + m_3 + is) \Gamma(1 - 2is) \Gamma(is - m_3 + 1/2)} \\
		&= \frac{i \pi}{2 s} \cdot \frac{\Gamma(1/2 - m_3 - is) \Gamma( 2is) \Gamma(-is + m_3 + 1/2)}
							{\Gamma(1/2 + m_3 + is) \Gamma( - 2is) \Gamma(is - m_3 + 1/2)}
\end{aligned}
\label{F_calculation_single_integral_1}
\end{equation}
Note that the pole at $s=0$ in this expression comes from the region $x \to 1$. 
According to \cite{LSZinLST}, this means that this pole is the \textquote{bulk pole}, that is, it does not signifies propagation of a physical state.

Putting all expressions together (and omitting the factor $\int d^{2} x|x|^{2\left(m_{1}+m_{2}+m_{3}-1\right)}$, which gives a conservation law for $m$'s) we arrive at \eqref{3pt_V-correlator_cont_j=0}:
\begin{equation}
	\left\langle  V_{\tilde{j}_{1} \to 0 ; m_{1}, - m_{1}} V_{-1/2 + is ; m_{2}, - m_{2}} V_{-1/2 - is ; m_{3}, - m_{3}} \right\rangle
	=
	\frac{1}{\tilde{j}_1}
	\times 	\frac{1}{2 \pi} e^{ i (\delta(s; m_2, -m_2 ; k) - \delta(s; m_3, -m_3 ; k)) }
\label{3pt_V-correlator_cont_j=0_appendix}	
\end{equation}

Finally we note that one can check the pole structure of the resulting correlation function \eqref{3pt_V-correlator_cont_j=0_appendix} by performing the calculation differently. Namely, we can first study the singularity at $x_1 \to 0$ in \eqref{3pt_GK_exact_Ffactor} which would give the pole at $\tilde{j}_1 + m_1 = -1,\, -2,\, \ldots$. The results are consistent.

One could worry that the pole at $\tilde{j} = 0$ might mix with a so called bulk pole (see the discussion in Sec.~2.4 in \cite{LSZinLST}). This can happen in the case when all but one vertex operators in a correlator are normalizable, but one operator has a normalizable and a non-normalizable part, both of which contribute to a pole. However in our case  the  operator with $\tilde{j}=0$ has  no bulk contribution, since
\begin{equation}
\begin{aligned}
	V_{\tilde{j}_1 \approx 0 ; m_{1}, -m_{1}} 
		&\sim e^{Q \cdot 0 \cdot \phi} + R(\tilde{j}_1\approx 0, m_1, -m_1 ; k) \cdot e^{- Q \phi} \\
		&\sim \text{Id} + R(\tilde{j}_1\approx 0, m_1, -m_1 ; k) \cdot e^{- Q \phi}
\end{aligned}
\label{fake_id_decomposition}
\end{equation}
where Id is the so-called \textquote{fake identity} operator, see e.g. \cite[Sec. 12.2.2]{Erbin}.
The second term here represents a normalizable operator times a pole.
Plugging \eqref{fake_id_decomposition} into the three point correlator we see that the second term reproduces the pole \eqref{3pt_V-correlator_cont_j=0_appendix}, while the first term with $\text{Id}$ gives the two-point function of two$j = -1/2 \pm is$ operators, which is finite and does not contain a pole.

\subsubsection{$j = -1/2$}
\label{sec:3pt_discr_calculation}

The three point function in the case with discrete representations $\tilde{j}_1 = 0$, $j_2 = j_3 = -1/2$  \eqref{3pt_V-correlator} can be obtained by formally sending $s$ to zero in the previous calculation. 
Setting $s=0$ in \eqref{3pt_V-correlator_cont_j=0} we recover \eqref{3pt_V-correlator}, \eqref{3pt_V-correlator_res}.

Moreover, careful examination of the limit $j_1 \to -1, j_2 \to -1/2, j_3 \to -1/2$ in \eqref{3pt_GK_exact} shows that the three-point function \eqref{3pt_1-half-half_normalizable} is indeed finite, cf. \eqref{3pt_V-correlator_res_with_normalizable}.
Setting $s=0$ in \eqref{3pt_V-correlator_cont_j=-1} we recover \eqref{3pt_V-correlator_res_with_normalizable}.

Another interesting question is the three point function with two logarithmic primary fields,
\begin{equation}
	\left\langle V_{\tilde{j}_1 = 0, m, -m} \ \phi V_{\tilde{j}_2 = -1/2, m, -m} \ \phi V_{\tilde{j}_3 = -1/2, m, -m} \right\rangle
\label{3pt_log_appendix}
\end{equation}
This three point function can be derived using the results of Appendix~\ref{sec:3pt_cont_calculation}.
To do this we recall that, first, in the non-supersymmetric Liouville theory we have
\begin{equation}
	\phi e^{ Q \phi} = -i \, \frac{\p}{\p s} \, e^{2 (\frac{Q}{2} + i s) \phi} \Bigg|_{s=0}
\end{equation}
and second that an analytic continuation of the three point function in the non-supersymmetric Liouville gives the structure constants
$D\left(\tilde{j}_1, j_2, j_3; k\right)$ \eqref{3pt_GK_exact_Dfactor} in the supersymmetric Liouville. 
From this we can conclude that we can obtain the three point function \eqref{3pt_log_appendix} as follows%
\footnote{There are also logarithmic terms coming from the world sheet dependence like in \eqref{nosusy_phi-e_2pt_via_3pt}, but these are subleading.}%
:
\begin{equation}
\begin{aligned}
	&\left\langle V_{\tilde{j}_1 = 0, m, -m} \ \phi V_{\tilde{j}_2 = -1/2, m, -m} \ \phi V_{\tilde{j}_3 = -1/2, m, -m} \right\rangle \\
	&= \Bigg\{ \left[  \frac{\p}{\p s_1} \frac{\p}{\p s_2}  D(0, -1/2+is_1, -1/2-is_2; k) \right] \\
	&\times F(0, m_1; -1/2+is_1, m_2; -1/2-is_2, m_3 )
	\int d^{2} x|x|^{2\left(m_{1}+m_{2}+m_{3}-1\right)}
	\Bigg\}
	\Bigg|_{s_1 = s_2 = 0}
\end{aligned}
\label{3pt_log_appendix_2}
\end{equation}
Performing the calculation, we obtain:
\begin{equation}
\begin{aligned}
	&\left\langle V_{\tilde{j}_1 = 0, m, -m} \ \phi V_{\tilde{j}_2 = -1/2, m, -m} \ \phi V_{\tilde{j}_3 = -1/2, m, -m} \right\rangle \\
	&= \frac{1}{2\pi} \frac{1}{ \tilde{j}_1 \, (j_2 + \frac{1}{2}) \, (j_3 + \frac{1}{2}) }
\end{aligned}
\label{3pt_log_appendix_3}
\end{equation}

%
%

\section{Comparison to the non-supersymmetric Liouville theory}
\label{sec:nosusy}

In this paper we mainly consider the SL($2, \mathbb{R}$)/U(1) coset WZNW model, which is a mirror of the \ntwo Liouville theory.
However it is interesting to compare with the results in non-supersymmetric Liouville theory.
From that we can draw lessons for our case (for example, why do we see a specific combination of exponentials in \eqref{GK_decomposition}) and find out some factorization properties.

Discussing various Liouville models below we will need to introduce the Liouville $b$-parameter. 
So in order not to confuse it with our VEV of the massless baryon in 4D SQCD, we will denote the Liouville parameter of the non-supersymmetric theory as $b_L$, and in the supersymmetric model as $b_{\mathcal{N}=2}$.
Moreover, we will denote the background charges in non-supersymmetric Liouville, supersymmetric Liouville, and SL($2, \mathbb{R}$)/U(1) coset WZNW respectively as $Q_L$, $Q_{\mathcal{N}=2}$ and $Q_{SL}$.

\subsection{Non-SUSY  vs. SUSY}

In the non-supersymmetric Liouville theory (considered e.g. in \cite{Dorn:1994xn,Zamolodchikov:1995aa,Harlow:2011ny}) the interaction term is
\begin{equation}
	\mathcal{L}_{int} \sim e^{2 b_L \phi_L}
\end{equation}
The requirement that this is a marginal deformation (of conformal weight 1) leads to
\begin{equation}
	b_L (Q_L - b_L) = 1 
	\Rightarrow
	Q_L = b_L + \frac{1}{b_L}
\end{equation}
see also \cite[eq. (2.13)]{Nakayama:2004vk}.


On the other hand, in the \ntwo Liouville the interaction term can be written as
\begin{equation}
	\mathcal{L}_{int} \sim \int d^2\theta e^{- \frac{1}{\sqrt{2}} b_{\mathcal{N}=2} (\phi_{\mathcal{N}=2} + i Y)}
\label{susy_b_interact}
\end{equation}
(Here we use the normalization where the asymptotic radius of the compact dimension is $R = \sqrt{2/k}$, see Sec.~\ref{sec:c=1}.)
In components we have terms like $b_{\mathcal{N}=2}^{2} \psi^{+} \psi^{-} e^{b_{\mathcal{N}=2} (\phi_{\mathcal{N}=2} + i Y)}$.
The requirement that this is a marginal deformation (of conformal weight 1) leads to
\begin{equation}
	\frac{1}{2} + \frac{b_{\mathcal{N}=2}}{2\sqrt{2}} ( Q_{\mathcal{N}=2} - \frac{b_{\mathcal{N}=2}}{2\sqrt{2}}) + \frac{b_{\mathcal{N}=2}^2}{8} = 1 
	\Rightarrow
	Q_{\mathcal{N}=2} = \frac{\sqrt{2}}{ b_{\mathcal{N}=2 } }
\label{Qn2_b}
\end{equation}
see also \cite[eq. (11.7)]{Nakayama:2004vk}. 
That is why \eqref{susy_b_interact} can be written as
\begin{equation}
	\mathcal{L}_{int} \sim \int d^2\theta e^{ - \frac{ \phi_{\mathcal{N}=2} + i Y}{Q_{\mathcal{N}=2 }} }
\label{susy_b_interact-2}
\end{equation}
cf. \cite[eq. (3.9)]{LSZinLST} and \cite[eq. (4.8)]{SYlittles}.

Next, consider the SL($2,\mathbb{R}$)/U(1) theory, which is a mirror to the \ntwo Liouville \cite{GVafa,GivKut,MukVafa,OoguriVafa95}.
In this model the asymptotic radius is $R=\sqrt{2k}$.
The  vertex operators at large positive $\phi$ are written as (see Sec.~\ref{sec:operators_2pt} for details)
\begin{equation}
	V \sim e^{Q_{SL} j \phi + Q_{SL} i m Y}
\end{equation}
%
Here, 
\begin{equation}
	Q_{SL} = Q_{\mathcal{N}=2} = \sqrt{\frac{2}{k}}
\label{Qsl=Qn2}
\end{equation}


\subsection{The dictionary}

Extending the dictionary from Teschner's papers \cite[eq. (17)]{Teschner:1999ug}, 
\cite[Sec. 4.4]{Teschner:1997ft} (see also \cite[eq. (6.92) and below]{Nakayama:2004vk})
we can write down the correspondence%
\footnote{In \cite{Ribault:2005wp,Chang:2014jta} an alternative dictionary was presented.}
between quantities in supersymmetric and non-supersymmetric Liouville theories:
\begin{equation}
\begin{aligned}
	b_L &\longleftrightarrow \frac{1}{\sqrt{k}} \\
	\alpha &\longleftrightarrow -   b_L  \, j \\
	Q_L &\longleftrightarrow \frac{1}{\sqrt{2}} Q_{SL},
\end{aligned}
\label{mydict-2}
\end{equation}
where $\alpha$ defines the primary operator $\exp{2\alpha \phi_L}$ in the Liouville theory.
(Also note that the level $k$ in Teschner's papers is shifted by 2.)
Comparing \eqref{Qn2_b}, \eqref{Qsl=Qn2}, \eqref{mydict-2} we conclude that $b_{\mathcal{N}=2 } = \frac{1}{b_L}$, and
\begin{equation}
	Q_{SL} = \sqrt{2} b_L 
\label{Qsl}
\end{equation}
This matches with  \cite[eq. (6.100)]{Nakayama:2004vk}, \cite[eq. (2.2)]{Iguri:2007af}.

For example, according to \eqref{mydict-2} and \eqref{Qsl}, the map acts as follows:
\begin{equation}
\begin{aligned}
	\alpha = Q_L / 2
	\longleftrightarrow
	j = - \frac{1}{b} \frac{1}{\sqrt{2}} Q_{SL} / 2 =  - 1 / 2
	\\
	Q_L - \alpha
	\longleftrightarrow
	- \frac{1}{b} (\frac{1}{\sqrt{2}}Q_{SL} + b_L j) = -1 -j
\end{aligned}
\end{equation}

Moreover, the Liouville field is mapped as
\begin{equation}
\begin{aligned}
	\phi_{L} = - \frac{1}{\sqrt{2}} \phi_{\mathcal{N}=2}
\end{aligned}
\label{mydict_phi-2}
\end{equation}
%

\subsection{Stress tensor}

The stress tensor is mapped correctly.
\begin{equation}
\begin{aligned}
	T_L &= -(\partial \phi_L)^{2}+Q_L \partial^{2} \phi_L \\
	&\to -(\partial \phi_L)^{2}+ \frac{1}{\sqrt{2}}Q_{SL} \partial^{2} \phi_L \\
	&=  - \frac{1}{2} \Bigg[ (\partial \phi_{\mathcal{N}=2})^{2} + \sqrt{\frac{2}{k}}  \partial^{2} \phi_{\mathcal{N}=2} \Bigg],  
\end{aligned}
\end{equation}
which is the $\phi$-part of the full stress tensor \eqref{T--}.
Primary operators (in the large-$\phi$ limit) and conformal dimensions are also mapped correctly: 
\begin{equation}
\begin{aligned}
	e^{2 \alpha \phi_L}
	\longleftrightarrow
	e^{2 (- \frac{j}{\sqrt{k}} ) (- \frac{1}{\sqrt{2}} \phi_{\mathcal{N}=2})  }
		= e^{\sqrt{2/k}j\phi_{\mathcal{N}=2}}
		= e^{Q_{SL}j \phi_{\mathcal{N}=2} }
	\\
	\Delta = \alpha (Q_L - \alpha)
	\longleftrightarrow 
	- b_L j (\frac{1}{\sqrt{2}}Q_{SL} + bj)
	= - b_L^2 j (j+1)
	 = - \frac{j (j+1)}{k}
\end{aligned}
\label{primary_mapping}
\end{equation}

Primary fields of the non-supersymmetric Liouville theory $\sim e^{2 \alpha \phi}$ can be both normalizable and non-normalizable.

\subsection{Reflection coefficient}

With this dictionary, the inverse%
\footnote{Some authors define the reflection coefficient $R$ (schematically) as $V_\alpha = R(\alpha) V_{Q - \alpha}$, while others use the definition $V_\alpha = R(Q - \alpha) V_{Q - \alpha}$. The two reflection coefficients are reciprocal with respect to each other. Our conventions in the bulk of this paper (and also the conventions of \cite{LSZinLST}) correspond to the former definition, while the authors of \cite{Zamolodchikov:1995aa,Harlow:2011ny} use the latter. \label{footnote:R_inverse}} 
reflection coefficient from \cite{Dorn:1994xn,Zamolodchikov:1995aa,Harlow:2011ny} correctly reproduces the $m$-independent part of the reflection coefficient \eqref{refl_GK}.
Indeed,
\begin{equation}
\begin{aligned}
	R(\alpha)_{L}
		= \left[\pi \mu_{L} \gamma\left(b_L^{2}\right)\right]^{(2 \alpha-Q_L) / b_L} 
			\frac{b_L^{2}}{\gamma\left(2 \alpha / b_L -1 -1 / b_L^{2}\right) \gamma\left(2 b_L \alpha-b_L^{2}\right)} \\
		= \left[\pi \mu_{L} \gamma\left(b_L^{2}\right)\right]^{(2 \alpha-Q_L) / b_L} 
			\frac{b_L^{2}}{\gamma\left(\frac{2 \alpha - Q_L}{b_L} \right) \gamma\left( b_L (2\alpha - Q_L) + 1 \right)} \\
		\longrightarrow \left[\pi \mu_{L} \gamma\left(b_L^{2}\right)\right]^{(2 (- b_L j) -Q_{SL}) / b_L} 
			\frac{b_L^{2}}{\gamma\left(\frac{2  (- b_L j)  - Q_{SL}}{b_L} \right) \gamma\left( b_L (2\alpha - Q_{SL}) + 1 \right)} \\
		= \left[\pi \mu_{L} \gamma\left( \frac{1}{k} \right)\right]^{ -2j-1 } 
			\frac{1}{k}
			\frac{1}{\gamma\left( -2j-1 \right) \gamma\left( 1 - \frac{2j+1}{k} \right)}
\end{aligned}
\end{equation}
Then, 
\begin{equation}
\begin{aligned}
	\frac{1}{R(\alpha)_{L}}
		\to k \left[\pi \mu_{L} \frac{\Gamma\left(\frac{1}{k}\right)}{\Gamma\left(1-\frac{1}{k}\right)} \right]^{ 2j+1 }
		\frac{\Gamma\left(1-\frac{2 j+1}{k}\right) \Gamma(-2 j-1)}{\Gamma\left(\frac{2 j+1}{k}\right)   \Gamma(2 j+2)} \\
		=  \left[k \pi \mu_{L} \frac{\Gamma\left(1+\frac{1}{k}\right)}{\Gamma\left(1-\frac{1}{k}\right)} \right]^{ 2j+1 }
		\frac{\Gamma\left(1-\frac{2 j+1}{k}\right) \Gamma(-2 j-1)}{\Gamma\left(1 + \frac{2 j+1}{k}\right)   \Gamma(2 j+1)} \\		
\end{aligned}
\label{refl_witten_mydict}
\end{equation}
On the other hand, the reflection coefficient in the supersymmetric Liouville theory is given by \eqref{refl_GK}.
One can see that that the analytically continued reflection coefficient in the non-supersymmetric Liouville \eqref{refl_witten_mydict} matches
the $m$-independent part of the supersymmetric reflection amplitude \eqref{refl_GK} up to the factor
$(k \pi^2 \mu_L)^{2j+1}$. (Recall that $m$ is the momentum along the compact direction, which is absent in the pure Liouville theory.)
However a factor of the form $(..)^{2j+1}$ is insignificant, since it can be absorbed into a redefinition of the operators \eqref{V_redef}.


\subsection{DOZZ formula}
\label{sec:DOZZ}

The formula for the three point correlation functionin non-supersymmetric pure Liouville theory (the so-called DOZZ formula) was derived in \cite{Dorn:1994xn,Zamolodchikov:1995aa}.
Later it was extensively studied by many authors; one can mention \cite{Harlow:2011ny,Teschner:1995yf,Kupiainen:2018snr} and a review \cite{Nakayama:2004vk}.
The structure constants $C\left(\alpha_{1}, \alpha_{2}, \alpha_{3}\right)$ (the analog of $D(j_1, j_2, j_3; k)$ from the supersymmetric case \eqref{3pt_GK_exact_Dfactor}) are
\begin{equation}
\begin{aligned}
	&C\left(\alpha_{1}, \alpha_{2}, \alpha_{3}\right)
	= \left[\pi \mu \gamma\left(b_L^{2}\right) b_L^{2-2 b_L^{2}}\right]^{\left(Q_L-\sum \alpha_{i}\right) / b_L}
	\\
	&\times
	\frac{\Upsilon_{0} \Upsilon\left(2 \alpha_{1}\right) \Upsilon\left(2 \alpha_{2}\right) \Upsilon\left(2 \alpha_{3}\right)}{\Upsilon\left(\alpha_{1}+\alpha_{2}+\alpha_{3}-Q_L\right) \Upsilon\left(\alpha_{1}+\alpha_{2}-\alpha_{3}\right) \Upsilon\left(\alpha_{2}+\alpha_{3}-\alpha_{1}\right) \Upsilon\left(\alpha_{1}+\alpha_{3}-\alpha_{2}\right)}
\end{aligned}
\label{3pt_no-susy}
\end{equation}
Here $Q_L = b_L + 1/b_L$.

Let us review the special function $\Upsilon$, see also \cite[eq.(3.10-13)]{Zamolodchikov:1995aa}, \cite[Appendix A]{Harlow:2011ny}.
It can be defined as
\begin{equation}
	\log \Upsilon(x)=\int_{0}^{\infty} \frac{d t}{t}\left[\left(\frac{Q_L}{2}-x\right)^{2} e^{-t}-\frac{\sinh ^{2}\left(\frac{Q_L}{2}-x\right) \frac{t}{2}}{\sinh \frac{b_L t}{2} \sinh \frac{t}{2 b_L}}\right]
\end{equation}
while
\begin{equation}
	\Upsilon_{0}=\left.\frac{d \Upsilon(x)}{d x}\right|_{x=0}
\end{equation}
%
Some useful properties of the $\Upsilon$-function are:
\begin{equation}
\begin{aligned}
	\Upsilon(x) &=\Upsilon(Q_L-x) \\
	\Upsilon(Q_L / 2) &=1
\end{aligned}
\end{equation}
\begin{equation}
\begin{aligned}
	\Upsilon(x+b_L) &=\gamma(b_L x) b_L^{1-2 b_L x} \Upsilon(x) \\
	\Upsilon(x+1 / b_L) &=\gamma(x / b_L) b^{2 x / b_L-1} \Upsilon(x)
\end{aligned}
\end{equation}
\begin{equation}
\begin{aligned}
	\Upsilon(x-b_L) &=\gamma\left(b_L x-b_L^{2}\right)^{-1} b_L^{2 b_L x-1-2 b_L^{2}} \Upsilon(x) \\
	\Upsilon(x-1 / b_L) &=\gamma\left(x / b_L-1 / b_L^{2}\right)^{-1} b_L^{1+\frac{2}{b_L^{2}}-\frac{2 x}{b_L}} \Upsilon(x)
\end{aligned}
\end{equation}
Here we use the standard notation 
$$\gamma(x) = \frac{\Gamma(x)}{\Gamma(1-x)}$$
The function $\Upsilon(x)$ has simple zeros at $x=-m b_L-n / b_L$ and $x = \left(m^{\prime}+1\right) b_L+\left(n^{\prime}+1\right) / b_L$. 
Here $n, n', m, m' =0,1,2, \ldots$.


\subsection{Different form of the structure constants in the supersymmetric Liouville theory}
\label{sec:3pt_susy}

The three point function in a corresponding WZNW model (which is a mirror of the \ntwo Liouville) was calculated in \cite{Teschner:1997ft,Teschner:1999ug}, see also \cite{Giveon:1999tq}. 
It is given by the product of the structure constants $D(j_1, j_2, j_3; k)$ (which do not depend on the momentum along the compact dimension $m$) and an $m$-dependent part.
It turns out that under the dictionary \eqref{mydict-2} the structure constants for this supersymmetric Liouville theory \eqref{3pt_GK_exact_Dfactor} maps to the three point function of the non-supersymmetric Liouville \eqref{3pt_no-susy}.

\subsubsection{Structure constants}

In \cite{Teschner:1997ft} the author considers the SL($2, \mathbb{C}$)/SU(2) WZNW model%
\footnote{The manifold SL($2, \mathbb{C}$)/SU(2) can be thought of as a Euclidean version of SL($2,\mathbb{R}$).}%
. The result for the three-point structure constants is \cite[eq. (64)]{Teschner:1997ft}
\begin{multline}
	C\left(\alpha_{1}, \alpha_{2}, \alpha_{3}\right) \\
	=
	\frac{C_{0}(b_L)(\nu(b_L))^{-b_L^{-1}\left(\alpha_{1}+\alpha_{2}+\alpha_{3}\right)} \Upsilon\left(2 \alpha_{1}\right) \Upsilon\left(2 \alpha_{2}\right) \Upsilon\left(2 \alpha_{3}\right)}
		{\Upsilon\left(\alpha_{1}+\alpha_{2}+\alpha_{3}-b_L \right) \Upsilon\left(\alpha_{1}+\alpha_{2}-\alpha_{3}\right) \Upsilon\left(\alpha_{1}+\alpha_{3}-\alpha_{2}\right) \Upsilon\left(\alpha_{2}+\alpha_{3}-\alpha_{1}\right)}
\label{3pt_Tech_1}
\end{multline}	
Here $C_{0}(b_L)(\nu(b_L))^{-b_L^{-1}\left(\alpha_{1}+\alpha_{2}+\alpha_{3}\right)}$ is a coefficient that can be found in 
\cite{Teschner:1997ft}.

Comparing this formula to the non-supersymmetric case \eqref{3pt_no-susy} we see that in the first $\Upsilon$-function in the denominator of \eqref{3pt_Tech_1} depends on  $b_L$ while in \eqref{3pt_no-susy} in the same place enters $Q_L$.
This fits with our dictionary \eqref{mydict-2}, \eqref{Qsl}. 
One can also compare to \cite[eq. (66)]{Teschner:1999ug}, \cite[eq. (23)]{Teschner:1999ug} (in the latter paper the author uses a different normalization of operators, see \cite[last paragraph of Appendix A]{Teschner:1999ug}).

The three point function formula can be also written in terms of the special $G$-function. 
Such form of the three point correlation functionis quite popular in the literature, so let us review it here, see also \cite[eq. (2.15-18)]{Maldacena:2001km}, \cite[eq. (A.3)]{Giveon:1999tq}, \cite[Appendix A]{LSZinLST}.
The function $G$ can be expressed via the Barnes double gamma function, see e.g. \cite[eq. (2.15)]{Maldacena:2001km}.
The relation between the $\Upsilon$ and the $G$ functions is given by
\begin{equation}
	G(j) = b_L^{-b_L^2 j (j+1+b_L^{-2})} \frac{1}{\Upsilon(b_L (j+1))}
\label{G_vs_Upsilon}
\end{equation}
cf. \cite[between eq. (23) and (24)]{Teschner:1999ug}%
\footnote{Note that Teschner uses in \cite{Teschner:1999ug} a different notation for $j$, namely, $j_{Teschner} = -j_{our}-1$.}%
.
Let us list some useful properties of the $G$-function:
\begin{equation}
	\begin{aligned}
	G(j) &=G(-j-1-k) \\
	G(j-1) &=\gamma\left(1+\frac{j}{k}\right) G(j) \\
	G(j-k) &=k^{-(2 j+1)} \gamma(j+1) G(j)
	\end{aligned}
\label{G_properties_GK}
\end{equation}
From \eqref{G_properties_GK} we can derive another useful property:
\begin{equation}
	\frac{G(j - 1)}{G(j)} \frac{G(- j - 1)}{G(- j)} = - \frac{j^2}{k^2} 
\label{G_property_1}
\end{equation}
The function $G(j)$ has poles at $j=n+m k$ and $j=-(n+1)-(m+1) k$, $n, m=0,1,2, \ldots$.

Using the relation \eqref{G_vs_Upsilon} and that $\alpha = - b_L j$ (see the dictionary \eqref{mydict-2}), one can easily see that the structure constants formulae \eqref{3pt_GK_exact_Dfactor} and \eqref{3pt_Tech_1} are equivalent.

\subsection{Factorization}

From what we have seen so far, we can make an observation.
The three point correlation function formula \eqref{3pt_GK_exact} for the SL($2,\mathbb{R}$)/U(1) (supersymmetric Liouville) theory can be naturally split into a product of \textquote{structure constants} that do not depend on the compact momentum $m$, and an $m$-dependent part. 
Moreover, it turns out that the structure constants in the supersymmetric theory \eqref{3pt_GK_exact_Dfactor}, \eqref{3pt_Tech_1}  precisely correspond to the analytically continued structure constants in the non-supersymmetric Liouville \eqref{3pt_no-susy}, when we use the dictionary \eqref{mydict-2}, \eqref{Qsl}.

So we see a curious property. 
The two and three point correlation functions in the supersymmetric Liouville factorize into the (analytically continued) \textquote{non-supersymmetric Liouville} part and an extra $m$-dependent part.
For the two point function it was noted in \cite{Teschner:1997ft}. In \cite{Ribault:2005wp} an alternative prescription was presented.

\subsection{Two point function from the DOZZ formula}
\label{sec:log_2pt}

In this section we are going to show that the two point function with non-normalizable $\tilde{j}=-1/2$ operators \eqref{V_j-1-half_nonnorm} 
\begin{equation}
	\langle \phi V_{\tilde{j} = -1/2, m, -m} \ \phi V_{\tilde{j} = -1/2, m, -m} \rangle \,.
\label{V_j-1-half_2pt_appendix}
\end{equation}
has a double pole.
To this end we are going to use the factorization property, see the previous subsection.
According to this property, the $m$-dependent part of the correlation function \eqref{V_j-1-half_2pt_appendix} is given simply by the $m$-dependent part of the reflection amplitude \eqref{refl_GK}, while for the $m$-independent part we can take the analytically continued result from the Liouville theory.

So let us consider for a moment the correlation function


\begin{equation}
	\langle \phi e^{2 \alpha \phi_L} \ \phi e^{ 2 \alpha \phi_L} \rangle \,,
	\quad
	\alpha = Q_L / 2 \,.
\label{phi-e_2pt}
\end{equation}
in the {\em non-supersymmetric} Liouville theory.
One way to compute this is to use the property%
\footnote{Note that $\p_j V_{j, m_L, m_R} \Big|_{j=-1/2}$ vanishes,
cf. \eqref{GK_decomposition} and
Sec.~\ref{sec:2pt_j=-1/2}.
}
 $\p_j e^{Q j \phi} \sim \phi \, e^{Q j \phi}$.
 
Let us start from the identity \cite{Dorn:1994xn,Zamolodchikov:1995aa,Harlow:2011ny} (see also Sec.~9.2.1 of \cite{Erbin})
\begin{equation}
	2\pi \, \delta(\alpha_1 - \alpha_2) \, \langle  e^{2 \alpha_1 \phi_L} \ e^{2 \alpha_2 \phi_L} \rangle 
		=  \underset{\varepsilon \to 0}{\lim} \langle  e^{ 2 \alpha_1 \phi_L} \ e^{ 2 \varepsilon \phi_L} \ e^{ 2 \alpha_2 \phi_L} \rangle \,.
\label{nosusy_e_2pt}
\end{equation}
The r.h.s. here can be computed explicitly using the formulas from Appendix~\ref{sec:DOZZ}.


Taking
\begin{equation}
	\alpha_1 = \frac{Q_L}{2} + i p_1 \,,
	\quad
	\alpha_2 = \frac{Q_L}{2} + i p_2 \,;
	\quad
	p_1 ,\, p_2 \to 0 \,;
	\quad
	\varepsilon \to 0
\end{equation}
we get (here we do write explicitly dependence on the world sheet coordinates):
%
\begin{equation}
\begin{aligned}
	\langle  e^{ 2 \alpha_1 \phi_L} \ e^{ 2 \varepsilon \phi_L} \ e^{ 2 \alpha_2 \phi_L} \rangle 
		&\approx  \frac{2\varepsilon}{\varepsilon^2 + (p_1 - p_2)^2} \, \frac{4 p_1 p_2}{\varepsilon^2 + (p_1 + p_2)^2} \\
			&\cross \frac{1}{|z_1 - z_2|^{2(\Delta_1 + \Delta_2 - \Delta_\varepsilon)} |z_1 - z_\varepsilon|^{2(\Delta_1 - \Delta_2 + \Delta_\varepsilon)} |z_2 - z_\varepsilon|^{2(- \Delta_1 + \Delta_2 + \Delta_\varepsilon)}} \\
		&\approx  2 \pi \delta(p_1 - p_2) \frac{(p_1 + p_2)^2}{\varepsilon^2 + (p_1 + p_2)^2} \\
			&\cross \frac{1}{|z_1 - z_2|^{2(\Delta_1 + \Delta_2 - \Delta_\varepsilon)} |z_1 - z_\varepsilon|^{2(\Delta_1 - \Delta_2 + \Delta_\varepsilon)} |z_2 - z_\varepsilon|^{2(- \Delta_1 + \Delta_2 + \Delta_\varepsilon)}} \,.
\end{aligned}
\label{nosusy_e_2pt_via_3pt}
\end{equation}
where $\Delta_i = \alpha_i (Q_L - \alpha_i) = Q_L^2/4 + p_i^2$, $\Delta_\varepsilon = \varepsilon (Q_L - \varepsilon)$, 
and $z_1$, $z_2$, $z_\varepsilon$ refer to the world sheet coordinates of the corresponding operators. 

To obtain the two point function \eqref{phi-e_2pt} we need to take derivatives of \eqref{nosusy_e_2pt_via_3pt} with respect%
\footnote{These are easiest to compute in terms of the variables $p_+ = p_1 + p_2$, $p_- = p_1 - p_1$.}
to $p_1$, $p_2$.
However this time we should take into account the dependence on the world sheet coordinates, since conformal dimensions depend on $p_1$, $p_2$. 
Taking the derivatives and dropping the $\delta'$, $\delta''$ terms, we obtain
%
\begin{equation}
\begin{aligned}
	\langle  \phi e^{ 2 \alpha_1 \phi_L} \ e^{ 2 \varepsilon \phi_L} \ \phi e^{ 2 \alpha_2 \phi_L} \rangle 
		&\approx  2 \pi \delta(p_1 - p_2) \frac{2 \varepsilon^2 (\varepsilon^2 - 12 p_1^2) }{(\varepsilon^2 + 4 p_1^2)^3} 
				\frac{1}{|z_1 - z_2|^{4 \Delta}} \\
		& 2 \pi \delta(p_1 - p_2) \frac{4 p_1^2}{\varepsilon^2 + 4 p_1^2} 
				\frac{\ln|z_1 - z_2|}{|z_1 - z_2|^{4 \Delta}} \,,
\end{aligned}
\label{nosusy_phi-e_2pt_via_3pt}
\end{equation}
The delta function lets us to set $p_1=p_2$.


When we send $\varepsilon \to 0$, the first term here generically vanishes. However, if $p_1$ and $p_2$ both tend to zero, the first term becomes singular. 
From this we see that the two point function \eqref{nosusy_e_2pt} in fact has a double pole, 
\begin{equation}
	\langle  \phi e^{- 2 \alpha_1 \phi} \ \phi e^{- 2 \alpha_2 \phi} \rangle 
		\sim \delta(\alpha_1 - \alpha_2) \frac{1}{(\alpha_1 - Q_L/2)^2} \,.
\label{nosusy_e_2pt_double-pole}
\end{equation}


From \eqref{nosusy_e_2pt_double-pole} and \eqref{borderline_2pt} we  conclude that the full correlation function \eqref{V_j-1-half_2pt_appendix} is given by
\begin{equation}
	\langle \phi V_{\tilde{j}_1 \to -1/2, m, -m} \ \phi V_{\tilde{j}_2 \to -1/2, m, -m} \rangle
		\sim \frac{1}{(\tilde{j}_1 + \frac{1}{2})^2} \delta(\tilde{j}_1 - \tilde{j}_2)
\label{V_j-1-half_2pt_result_appendix}
\end{equation}

\subsection{Pure Liouville degenerate operators}
\label{sec:pure_degenerate}

One may wonder why in the primary operators \eqref{GK_decomposition} we see only the specific combination 
$$e^{Q_{\mathcal{N}=2} j \phi_{\mathcal{N}=2}}+ R(j, m_L, m_R ; k) e^{-Q_{\mathcal{N}=2}(j+1) \phi_{\mathcal{N}=2}}$$
while the orthogonal combination (with the minis sign in front of $R$) does not appear. 
We can see the hint of this in non-supersymmetric Liouville theory.

The reflection property in the non-supersymmetric Liouville theory \cite{Dorn:1994xn,Zamolodchikov:1995aa} can be written as
\begin{equation}
	e^{2(Q_L-\alpha) \phi_L} = R(\alpha)_L e^{2 \alpha \phi_L}
\label{reflection_exp}
\end{equation}
(See the footnote \ref{footnote:R_inverse} on page \pageref{footnote:R_inverse}.)
The two-point function is structurally similar to \eqref{2pt_via_refl},
\begin{equation}
	\langle e^{2 \alpha \phi_L} e^{2 \alpha \phi_L} \rangle \sim \frac{1}{R(\alpha)_L} 
\label{2pt_witten}	
\end{equation}

The formula \eqref{reflection_exp} should be interpreted as a valid relation on the level of correlation functions.
For example, using \eqref{reflection_exp} and \eqref{2pt_witten} one can see that
\begin{equation}
	\langle e^{2 (Q-\alpha) \phi} e^{2 \alpha \phi} \rangle
		=  R(\alpha) \langle e^{2 \alpha \phi} e^{2 \alpha \phi} \rangle
		= 1
\label{2pt_witten_reflected}
\end{equation}
Moreover,
\begin{equation}
	\langle (e^{2(Q_L-\alpha) \phi_L} - R(\alpha)_L e^{2 \alpha \phi_L}) \cdot (e^{2(Q_L-\alpha) \phi_L} - R(\alpha)_L e^{2 \alpha \phi_L}) \rangle = 0
\end{equation}
Therefore, the combination $e^{2(Q_L-\alpha) \phi_L} - R(\alpha)_L e^{2 \alpha \phi_L}$ is in fact a null-state,  and in this sense we can understand \eqref{reflection_exp}.
Of course, the reflection coefficient $R(\alpha)_L$ is different from \eqref{refl_GK}, but they are closely related, see 
Sec. B.4.

%
%

\section{Two point correlation function check}
\label{sec:2pt_check}

In this Appendix we use a quasiclassical approximation  to check the consistency of the two point function formula \eqref{2pt_via_refl}. 
Looking at \eqref{GK_decomposition}, \eqref{V_non-normalizable} and \eqref{2pt_via_refl} together one may wonder whether these  formulas are consistent. Namely,  we would like to compare the pole structure of r.h.s and l.h.s of \eqref{2pt_via_refl}.

Consider a correlation function of two non-normalizable operators.
Suppose we use the expansion \eqref{GK_decomposition} in the two-point function \eqref{2pt_via_refl} and go at one of the poles of $R(j, m, -m ; k)$.
Then it seems that in the r.h.s. of \eqref{2pt_via_refl} we get a single pole, which is fine for a two-point function. However, according to \eqref{GK_decomposition} and \eqref{V_non-normalizable},
on the l.h.s. both operators develop a pole%
\footnote{It is important that  on the one hand \eqref{discrete} is symmetric in $m \leftrightarrow -m$, and on the other hand $R(j, m, -m ; k)$ has the same poles as $R(j, -m, m ; k)$.}%
, and it seems that we should get a double pole. 
What's going on?

To resolve this issue it is easier to make the calculation of the correlation function in the Euclidean AdS$_3$ theory with primary operators $\Phi_{j ; m_L, m_R}$ (it is equal to the correlation function of the coset operators, see e.g. \cite[eq. (3.4)]{Giveon:1999tq}).
Large $\phi$ expansion of these operators is similar to \eqref{2pt_via_refl}:
\begin{equation}
	\Phi_{j ; m_L, m_R}
		\approx   e^{Q j \phi} \gamma^{j+m_L} \bar{\gamma}^{j - m_R}
		+   R(j, m_L, m_R ; k) \cdot e^{-Q(j+1) \phi} \gamma^{m_L-j-1} \bar{\gamma}^{- m_R -j-1}
\label{large_phi}
\end{equation}
Here, $(\phi, \gamma, \bar{\gamma})$ are the Poincar{\'e} coordinates on Euclidean AdS$_3$ \cite[eq. (2.14)]{LSZinLST}.
The two point function of these operators is the same as for the $V$-operators, see e.g. \cite[eq. (3.4), (3.6)]{Giveon:1999tq}:
\begin{equation}
	\left\langle\Phi_{j_1 ; m_L, m_R} \Phi_{j_2 ; -m_L, -m_R}\right\rangle
	=
	R(j_1, m_L, m_R ; k) \,
	\delta(j_1 - j_2)
\label{2pt_Phi}
\end{equation}

Let's concentrate on the case of interest 
$ m_L = -m_R \equiv m $.
Recall that the reflection coefficient $R(j, m, -m ; k)$ \eqref{refl_GK} has poles at special values of $j, m$ corresponding to discrete representations \eqref{discrete}.


When calculating the correlation function of two operators \eqref{large_phi}, we have to integrate over the (zero modes of) $\gamma$'s. Moreover, the $\gamma$'s in the first operator $\Phi_{j ; m, -m}$ and in the second operator $\Phi_{j ;-m, m}$ in \eqref{2pt_Phi} are basically the same, since $\gamma(z) \gamma(z') \sim \gamma^2(z)(1 + O(z-z'))$ (here $z,z'$ are the world sheet coordinates).

Then, from \eqref{large_phi} we have:
\begin{equation}
\begin{aligned}
	\left\langle\Phi_{j_1 ; m, -m} \Phi_{j_2 ; -m, m}\right\rangle &\sim 
		  \int d^2 \gamma \Big\{
		\langle e^{Q j_1} e^{Q j_2} \rangle |\gamma|^{2(j_1 + j_2)} \\
		&+  \langle e^{Q j_1} e^{-Q (j_2+1)} \rangle |\gamma|^{-2} 
				R(j_2, -m, m ; k)  \\
		&+  \langle e^{-Q (j_1+1)} e^{Q j_2} \rangle |\gamma|^{-2} 
				R(j_1, m, -m ; k)  \\
		&+ \langle e^{-Q (j_1+1)} e^{-Q (j_2+1)} \rangle |\gamma|^{-2(j_1 + j_2+2)} \\
			&\phantom{+ee} \times  R(j_1, m, -m ; k)  R(j_2, -m, m ; k) 
		\Big\}
\end{aligned}
\label{2pt_Phi_expanded_1}
\end{equation}
Now we need to integrate over $\gamma$. Using the integration formula \eqref{master_integral} we obtain
\begin{equation}
	\int\limits_{\mathbb{C}} |\gamma|^{-2a} d^2 \gamma = \pi \frac{\Gamma(1-a) \Gamma(1) \Gamma(a-1)}{\Gamma(a) \Gamma(0) \Gamma(2-a)}
\end{equation}
But this is zero, unless $a = 1$! The divergence at $a = 1$ can be interpreted as the volume of the target space, cf. e.g. 
\cite[below eq. (3.5)]{GivKut}, \cite[eq. (5.12)]{Maldacena:2001km}.

From this we see that there are two cases: either $j_1 + j_2 = -1$ or $j_1 + j_2 \neq -1$.
Let us start with the latter.
Integrating over $\gamma$ in \eqref{2pt_Phi_expanded_1} we obtain
\begin{equation}
\begin{aligned}
	\left\langle\Phi_{j_1 ; m, -m} \Phi_{j_2 ;-m,m}\right\rangle \sim 
			&\langle e^{Q j_1} e^{-Q (j_2+1)} \rangle 
					R(j_2, -m, m ; k)  \\
			&+  \langle e^{-Q (j_1+1)} e^{Q j_2} \rangle 
					R(j_1, m, -m ; k) 
\end{aligned}
\label{2pt_Phi_expanded_2}
\end{equation}
(The divergent $\gamma$-integral cancels in the SL($2, \mathbb{R}$)/U(1) coset theory, see \cite[Sec. 3.2]{Chang:2014jta}.)
Thus, the last term of \eqref{2pt_Phi_expanded_1} (containing a would-be double pole) drops out.
Now it's time to say something  about correlation functions of pure exponentials. 
They do not contain any $m$- or $\gamma$-dependence.
Therefore these correlation functions can be computed using the results from non-supersymmetric Liouville theory, see Appendix~\ref{sec:nosusy}.
From this logic we obtain (cf. \eqref{2pt_witten_reflected})
\begin{equation}
	\langle e^{-Q (j_1+1)} e^{Q j_2} \rangle = \delta(j_1 - j_2) 
	\quad
	\text{in the vicinity  }  j_1 \approx j_2
\end{equation}
Using this in \eqref{2pt_Phi_expanded_2} we arrive at
\begin{equation}
	\left\langle\Phi_{j_1 ; m, -m} \Phi_{j_2 ;m,-m}\right\rangle \sim
		\delta(j_1 - j_2)
		R(j_1, m, -m ; k)
\end{equation}
This has the same pole structure as the formula \eqref{2pt_Phi}.

To conclude we comment on the case $j_1 = j_2 = -1/2$. 
In this case, all the terms in \eqref{2pt_Phi_expanded_1} do contribute. 
However, they all give comparable contributions, since the reflection coefficient does not actually develop a pole. 
This is consistent with the results of Sec.~\ref{sec:2pt_j=-1/2}.


%
%

\section{Useful formulas}
\label{sec:useful}

Integration formula useful for the $m$-mode expansion of SL($2, \mathbb{R}$) primary operators 
(see e.g. \cite{Giveon:1999tq,Maldacena:2001km}):
\begin{equation}
	\int\limits_{\mathbb{C}} d^{2} x|x|^{2 a} x^{n}|1-x|^{2 b}(1-x)^{m}
	=\pi \frac{\Gamma(a+n+1) \Gamma(b+m+1) \Gamma(-a-b-1)}{\Gamma(-a) \Gamma(-b) \Gamma(a+b+m+n+2)}
	, \quad n, m \in \mathbb{Z}
\label{master_integral}
\end{equation}
This formula can be derived using the method outlined in the book by Green, Schwarz, Witten \cite{GSW} (Sec. 7.2.2-7.2.3 there).
When the integral in the l.h.s. does not converge, we will use the r.h.s. as the analytic continuation.

Another useful formula that helps with derivations of Sec. \ref{sec:3pt_VVV}:
integrating by parts, one can show that
\begin{equation}
	\underset{\epsilon \to 0}{\Res} \int\limits_0^\infty dx \, x^{-1 + \epsilon} \cdot f(x) = f(0)
\end{equation}
at least if $f$ is differentiable and $f(x\to\infty) = 0$. (However, this result seem to hold even if $f$ has singularities and is divergent at the infinity.)

The residue \eqref{j=-1_res} can be calculated as follows.
Consider $\tilde{j} = \epsilon \ll 0$, and take half-integer $m > 0$. We have:
\begin{equation}
\begin{aligned}
	R&(\tilde{j} = \epsilon, m, -m ; k) \\
		&= \left[ \frac{1}{\pi} \frac{\Gamma\left(1+\frac{1}{k}\right)}{\Gamma\left(1-\frac{1}{k}\right)} \right]^{2\epsilon+1}
				\frac{\Gamma\left(1-\frac{2 \epsilon+1}{k}\right) \Gamma(m+\epsilon+1) \Gamma(-m+\epsilon+1) \Gamma(-2 \epsilon-1)}%
				{\Gamma\left(1+\frac{2 \epsilon+1}{k}\right) \Gamma(m-\epsilon) \Gamma(-m-\epsilon) \Gamma(2 \epsilon+1)} \\
		&\approx \frac{1}{\pi}  \left[ \frac{1}{\pi} \frac{\Gamma\left(1+\frac{1}{k}\right)}{\Gamma\left(1-\frac{1}{k}\right)} \right]^{2\epsilon} 
						\frac{  \Gamma(m+1) \Gamma(-m+\epsilon+1) \Gamma(-2 \epsilon-1)}%
						{ \Gamma(m) \Gamma(-m-\epsilon) \Gamma(1)} \\
		&\approx \frac{1}{\pi} m \cdot \frac{   \Gamma(-m+\epsilon+1) \Gamma(-2 \epsilon-1)}%
							{  \Gamma(-m-\epsilon) }
\end{aligned}
\end{equation}
Next we use the formula
\begin{equation}
	\Gamma (-n + \epsilon) \approx \frac{(-1)^n}{\epsilon \, n!}
\end{equation}
Then,
\begin{equation}
\begin{aligned}
	R(\tilde{j} = \epsilon, m, -m ; k) 
		&\approx \frac{1}{\pi} m \cdot \frac{   \Gamma(-m+\epsilon+1) \Gamma(-2 \epsilon-1)}%
							{  \Gamma(-m-\epsilon) } \\
		&\approx \frac{1}{\pi} m \cdot \frac{(-1)^{m-1}}{\epsilon  (m-1)!} \cdot \frac{(-1)^1}{(-2\epsilon)  (1)!} \cdot  \frac{(-\epsilon)  m!}{(-1)^m} \\
		&= \frac{m^2}{2 \pi} \cdot \frac{1}{\epsilon}
\end{aligned}
\end{equation}

Finally, let us also note  a typo in Ref.~\cite{Giveon:1999tq}. 
Using the formulae from this Appendix and from Appendix~\ref{sec:3pt_additional} one can check that factorial signs in the numerator in the last line of eq. (4.20) in \cite{Giveon:1999tq} should not be there. So, the formula (4.20) in \cite{Giveon:1999tq} should read
\begin{equation}
\begin{aligned}
	S(2 ; 3)=\sum_{n=\max \left\{0, N+1-n_{2}\right\}}^{\min \left\{N, n_{3}-1\right\}}
	{N \choose n}
	\frac{(-1)^{n_{3}-1-n}}{\left(n_{3}-1-n\right) !\left(n_{2}-1+n-N\right) !} \times
	\\
	\prod_{i=0}^{n_{3}-n-2}\left(2 j_{3}+n_{3}+N-n-i\right)  \prod_{i=0}^{n_{2}+n-N-2}\left(2 j_{2}+n_{2}+n-i\right) 
	\,.
\end{aligned}
\end{equation}

\end{appendices}

%
%


\clearpage



\begin{thebibliography}{99}
\addcontentsline{toc}{section}{References}
\itemsep -2pt



\bibitem{HT1}
A.~Hanany and D.~Tong,
{\em Vortices, instantons and branes,}
JHEP {\bf 0307}, 037 (2003).
[hep-th/0306150].

\bibitem{ABEKY}
R.~Auzzi, S.~Bolognesi, J.~Evslin, K.~Konishi and A.~Yung,
{\em Non-Abelian superconductors: Vortices and
 confinement in ${\mathcal N}=2$  SQCD,}
Nucl.\ Phys.\ B {\bf 673}, 187 (2003).
[hep-th/0307287].

\bibitem{SYmon}
M.~Shifman and A.~Yung,
{\em Non-Abelian string junctions as confined monopoles,}
Phys.\ Rev.\ D {\bf 70}, 045004 (2004)
[hep-th/0403149].

 \bibitem{HT2}
A. Hanany and D. Tong,
{\em Vortex strings and four-dimensional gauge dynamics,}
JHEP {\bf 0404}, 066 (2004)
[hep-th/0403158].



\bibitem{ANO} 
A.~Abrikosov,
{\em On the Magnetic Properties of Superconductors of the Second Group,}
 Sov.~Phys. JETP {\bf5}, 1174  (1957);
 Russian original -- ZhETF {\bf32}, 1442  (1957);\\
H.~Nielsen and P.~Olesen, 
{\em Vortex-line models for dual strings,}
Nucl.~Phys. {\bf B61}, 45 (1973).
[Reprinted in {\em Solitons and Particles}, Eds. C. Rebbi and G. Soliani
(World Scientific, Singapore, 1984), p. 365].


\bibitem{Trev}
D.~Tong, {\em TASI Lectures on Solitons,}
  arXiv:hep-th/0509216.

\bibitem{Jrev}
  M.~Eto, Y.~Isozumi, M.~Nitta, K.~Ohashi and N.~Sakai,
{\em Solitons in the Higgs phase: The moduli matrix approach,}
  J.\ Phys.\ A  {\bf 39}, R315 (2006)
  [arXiv:hep-th/0602170].
  
\bibitem{SYrev}
M.~Shifman and A.~Yung,
{\em Supersymmetric Solitons and How They Help Us Understand Non-Abelian Gauge Theories,}
  Rev.\ Mod.\ Phys.\  {\bf 79}, 1139 (2007)
  [hep-th/0703267]; for an expanded version see
{\sl Supersymmetric Solitons,}
(Cambridge University Press, 2009).

\bibitem{Trev2}
D.~Tong,
{\em Quantum Vortex Strings: A Review,}
  Annals Phys.\  {\bf 324}, 30 (2009)
  [arXiv:0809.5060 [hep-th]].
	
\bibitem{SYcstring} 
  M.~Shifman and A.~Yung,
{\em Critical String from Non-Abelian Vortex in Four Dimensions,}
  Phys.\ Lett.\ B {\bf 750}, 416 (2015)
  [arXiv:1502.00683 [hep-th]].

  \bibitem{FI}
  P.~Fayet and J.~Iliopoulos,
{\em Spontaneously Broken Supergauge Symmetries and Goldstone Spinors,}
  Phys.\ Lett.\  B {\bf 51}, 461 (1974).

	 \bibitem{Candel}
P.~Candelas and X.~C.~ de la Ossa,
{\em Comments on conifolds,}
Nucl. \ Phys. \ {\bf B342}, 246 (1990).

\bibitem{NVafa}
A.~Neitzke and  C.~Vafa,
{\em Topological strings and their physical applications},
arXiv:hep-th/0410178.
	
 \bibitem{KSYconifold}
P.~Koroteev, M.~Shifman and A.~Yung,
{\em  Non-Abelian Vortex in
 Four Dimensions as a Critical  String on a Conifold},
 Phys.\ Rev.\ D {\bf 94} (2016) no.6,  065002
  [arXiv:1605.08433 [hep-th]].

\bibitem{KSYcstring}
P.~Koroteev, M.~Shifman and A.~Yung,
{\em Studying Critical  String Emerging from 
Non-Abelian Vortex in Four Dimensions},
Phys. \ Lett. \ {\bf B759}, 154 (2016)  
[arXiv:1605.01472 [hep-th]].
 
\bibitem{Ievlev:2020qch}
E.~Ievlev, M.~Shifman and A.~Yung,
{\em String baryon in four-dimensional $\mathcal{N}$=2 supersymmetric QCD from the 2D-4D correspondence,}
Phys. Rev. D \textbf{102}, no.5, 054026 (2020)
[arXiv:2006.12054 [hep-th]].



\bibitem{SYlittles} 
  M.~Shifman and A.~Yung,
{\em Critical Non-Abelian Vortex in Four Dimensions and Little String Theory,}
  Phys.\ Rev.\ D {\bf 96}, no. 4, 046009 (2017)
  [arXiv:1704.00825 [hep-th]].


\bibitem{SYlittmult} 
 M.~Shifman and A.~Yung,
 {\em Hadrons of $\mathcal N=2$ Supersymmetric QCD in Four Dimensions from Little String Theory,}
  Phys.\ Rev.\ D {\bf 98}, no. 8, 085013 (2018)
  [arXiv:1805.10989 [hep-th]].
  
  



\bibitem{Kutasov}
D.~Kutasov,
{\em Introduction to Little String Theory}, published in
{\sl Superstrings and Related Matters 2001},  Proc. of the ICTP Spring School 
of Physics, Eds. C. Bachas, K.S. Narain, and   S. Randjbar-Daemi, 2002,  pp.165-209.
  
\bibitem{GVafa}
D.~Ghoshal and  C.~Vafa, 
{\em c = 1 String as the Topological Theory of the Conifold},
Nucl.\ Phys.\ B {\bf 453}, 121 (1995)
  [hep-th/9506122].
	
\bibitem{GivKut}
A.~Giveon and D.~Kutasov,
{\em Little String Theory in a Double Scaling Limit},
JHEP {\bf 9910}, 034 (1999)
  [hep-th/9909110].
	
  \bibitem{GivKutP}
A.~Giveon, D.~Kutasov and O.~Pelc,
{\em Holography for Noncritical Superstrings},
JHEP {\bf 9910}, 035 (1999)
  [hep-th/9907178].
  
  
\bibitem{HoriKapustin}
	 K.~Hori and A.~Kapustin,
{\em Duality of the fermionic 2-D black hole and N=2 Liouville theory as mirror symmetry,}
  JHEP {\bf 0108}, 045 (2001)
  [hep-th/0104202].


\bibitem{Wbh}
E.~Witten,
{\em String Theory and Black Holes,}
Phys.\ Rev.\ D {\bf 44}, 314 (1991).
  
\bibitem{MukVafa}
S.~Mukhi and C.~Vafa,
{\em Two-dimensional black hole as a topological coset model of c = 1 string theory},
Nucl. \ Phys.\ {\bf B407}  667, (1993)
[arXiv: hep-th/9301083].

  \bibitem{OoguriVafa95}
H.~Ooguri and C.~Vafa,
{\em Two-Dimensional Black Hole and Singularities of CY Manifolds,}
Nucl.\ Phys.\ B {\bf 463}, 55 (1996)
  [hep-th/9511164].
  
  	
\bibitem{DixonPeskinLy}
 L.~J.~Dixon, M.~E.~Peskin and J.~D.~Lykken,
{\em N=2 Superconformal Symmetry and SO(2,1) Current Algebra,}
  Nucl.\ Phys.\ B {\bf 325}, 329 (1989).

\bibitem{Petrop}
P.M.S.~Petropoulos,
{\em Comments on SU(1,1) string theory},
Phys. \ Lett.\ {\bf B236}, 151 (1990).

\bibitem{Hwang}
S.~Hwang,
{\em Cosets as Gauge Slices in SU(1,1) Strings,}
Phys. \  Lett.\ {\bf B276} 451, (1992) 
[arXiv:hep-th/9110039].

\bibitem{EGPerry}
J.~M.~Evans, M.~R.~Gaberdiel and M.~J.~Perry,
{\em The no ghost theorem for AdS(3) and the stringy exclusion principle,}
  Nucl.\ Phys.\ B {\bf 535}, 152 (1998)
  [hep-th/9806024].

\bibitem{Dorn:1994xn}
H.~Dorn and H.~J.~Otto,
{\em Two and three point functions in Liouville theory,}
Nucl. Phys. B \textbf{429}, 375-388 (1994)
[arXiv:hep-th/9403141 [hep-th]].


\bibitem{Zamolodchikov:1995aa}
A.~B.~Zamolodchikov and A.~B.~Zamolodchikov,
{\em Structure constants and conformal bootstrap in Liouville field theory,}
Nucl. Phys. B \textbf{477}, 577-605 (1996)
[arXiv:hep-th/9506136 [hep-th]].



\bibitem{Teschner:1997ft}
J.~Teschner,
{\em On structure constants and fusion rules in the SL($2, \mathbb{C}$) / SU(2) WZNW model,}
Nucl. Phys. B \textbf{546}, 390-422 (1999)
[arXiv:hep-th/9712256 [hep-th]].


\bibitem{Teschner:1999ug}
J.~Teschner,
{\em Operator product expansion and factorization in the H+(3) WZNW model,}
Nucl. Phys. B \textbf{571}, 555-582 (2000)
[arXiv:hep-th/9906215 [hep-th]].


\bibitem{Dijkgraaf:1991ba}
R.~Dijkgraaf, H.~L.~Verlinde and E.~P.~Verlinde,
{\em String propagation in a black hole geometry,}
Nucl. Phys. B \textbf{371}, 269-314 (1992)


\bibitem{Nakayama:2004vk}
Y.~Nakayama,
{\em Liouville field theory: A Decade after the revolution,}
Int. J. Mod. Phys. A \textbf{19}, 2771-2930 (2004)
[arXiv:hep-th/0402009 [hep-th]].




\bibitem{AharGubserMaldaOoguriOz}
O.~Aharony, S.~Gubser, J.~Maldacena, H.~Ooguri  and Y.~Oz,
{\em Large N Field Theories, String Theory and Gravity},
 Phys. \ Rept. \textbf{323}, 183 (2000) 
[hep-th/9905111]

\bibitem{ABKS}
O.~Aharony, M.~Berkooz, D.~Kutasov and  N.~Seiberg, 
{\em Linear dilatons, NS five-Branes and Holography},
JHEP {\bf 9810}, 004 (1998)
  [hep-th/9808149].
  
\bibitem{LSZinLST}
O.~Aharony, A.~Giveon and D.~Kutasov,
{\em LSZ in LST,}
Nucl. Phys. B \textbf{691}, 3-78 (2004)
[arXiv:hep-th/0404016 [hep-th]].




	\bibitem{T}
D.~Tong,
{\em Monopoles in the Higgs phase,}
Phys.\ Rev.\ D {\bf 69}, 065003 (2004)
[hep-th/0307302].

	
\bibitem{SYdualrev}
M.~Shifman and A.~Yung,
{\em Lessons from supersymmetry: ``Instead-of-Confinement'' Mechanism,}
  Int.\ J.\ Mod.\ Phys.\ A {\bf 29}, no. 27, 1430064 (2014)
  [arXiv:1410.2900 [hep-th]].
    
\bibitem{SW2} 
N.~Seiberg and E.~Witten,
{\em Monopoles, duality and chiral symmetry breaking in N=2 supersymmetric QCD,}
Nucl. Phys. {\bf B431}, 484  (1994)
[hep-th/9408099].

\bibitem{APS}
P.~Argyres, M.~Plesser and N.~Seiberg,
{\em The Moduli Space of ${\mathcal N}=2$  SUSY QCD and Duality in
${\mathcal N}=1$  SUSY QCD,}
Nucl. Phys. {\bf B471}, 159  (1996)
 [hep-th/9603042].
  
	\bibitem{AchVas}
For a review see e.g. A.~Achucarro and T.~Vachaspati,
{\em Semilocal and electroweak strings,}
  Phys.\ Rept.\  {\bf 327}, 347 (2000)
  [hep-ph/9904229].
  
\bibitem{SYsem}
 M.~Shifman and A.~Yung,
  {\em Non-Abelian semilocal strings in  ${\mathcal N} = 2$ supersymmetric QCD,}
  Phys.\ Rev.\  D {\bf 73}, 125012 (2006)
  [arXiv:hep-th/0603134].
  
 \bibitem{Jsem}
M.~Eto, J.~Evslin, K.~Konishi, G.~Marmorini, et al.,
{\em On the moduli space of semilocal strings and lumps,}
  Phys.\ Rev.\  D {\bf 76}, 105002 (2007)
  [arXiv:0704.2218 [hep-th]].
  
\bibitem{SVY}
 M.~Shifman, W.~Vinci and A.~Yung,
{\em Effective World-Sheet Theory for Non-Abelian Semilocal 
Strings in ${\mathcal N} = 2$ Supersymmetric QCD,}
  Phys.\ Rev.\ D {\bf 83}, 125017 (2011)
  [arXiv:1104.2077 [hep-th]].
	
	  \bibitem{Gepner}
D.~Gepner,
{\em Space-Time Supersymmetry in Compactified String Theory and Superconformal Models,}
Nucl.\ Phys.\ B {\bf 296}, 757 (1988).

\bibitem{BDFM}
T.~Banks, L.~J.~Dixon, D.~Friedan, and E.~J.~Martinec, 
{\em Phenomenology and Conformal Field Theory Or Can String Theory Predict  the Weak Mixing Angle?}
Nucl. \ Phys. \ B  {\bf 299}, 613 (1988).
	
	
\bibitem{ArgPlessShapiro}
P.~Argyres, M.~R.~Plesser and  A.~Shapere,
{\em The Coulomb Phase of ${\cal N}=2$ Supersymmetric QCD}
Phys.\ Rev. \ Lett. {\bf 75}, 1699 (1995)
[hep-th/9505100].

\bibitem{Chandrasekharan}
K.~Chandrasekharan,
{\sl Elliptic Functions,}
(Springer-Verlag, 1985).
DOI: 10.1007/978-3-642-52244-4.



\bibitem{Dorey}
N.~Dorey,
{\em The BPS spectra of two-dimensional supersymmetric gauge theories with  twisted mass terms,}
JHEP {\bf 9811}, 005 (1998) [hep-th/9806056].

\bibitem{DoHoTo}
N.~Dorey, T.~J.~Hollowood and D.~Tong,
{\em The BPS spectra of gauge theories in two and four dimensions,}
  JHEP {\bf 9905}, 006 (1999)
  [arXiv:hep-th/9902134].
  
  \bibitem{Karasik}
  E. Gerchkovitz and A. Karasik, {\em New Vortex-String World-sheet Theories from Super-Symmetric Localization}, 
  arXiv:1711.03561 [hep-th].
 
\bibitem{EGPerry-rev}
J.M.~Evans, M.R.~Gaberdiel and M.J~Perry,
{\em The no-ghost theorem and strings on AdS$_3$},
[hep-th/9812252], published in Proc. 1998 ICTP Spring School of Physics
  {\sl Nonperturbative Aspects of Strings, Branes and Supersymmetry},  Eds. M. J. Duff {\em et al.},
  pp. 435-444.
   
	 
\bibitem{KutSeib}
D.~Kutasov and N.~Seiberg,
{\em Non-critical Superstrings,}
Phys.\ Lett.\ B {\bf 251}, 67 (1990).
	
\bibitem{YNSflux}
A.~Yung,
{\em Flux Compactification for the Critical Non-Abelian Vortex and Quark Masses}
Phys.\ Rev.\ D {\bf 104}, 025007 (2021)
[arXiv:2105.02645 [hep-th]].


\bibitem{KlebWitten}
I.~R.~Klebanov and E.~Witten, 
{\em Superconformal Field Theory on Threebranes at a Calabi-Yau Singularity,}
 Nucl. \ Phys.  {\bf B536}, 199 (1998), 
[hep-th/9807080].

\bibitem{Giveon:1999tq}
A.~Giveon and D.~Kutasov,
{\em Comments on double scaled little string theory,}
JHEP \textbf{01}, 023 (2000)
[arXiv:hep-th/9911039 [hep-th]].


\bibitem{Erbin:2019uiz}
H.~Erbin, J.~Maldacena and D.~Skliros,
{\em Two-Point String Amplitudes,}
JHEP \textbf{07}, 139 (2019)
doi:10.1007/JHEP07(2019)139
[arXiv:1906.06051 [hep-th]].



\bibitem{Maldacena:2000hw}
J.~M.~Maldacena and H.~Ooguri,
{\em Strings in AdS(3) and SL($2, \mathbb{R}$) WZW model 1.: The Spectrum,}
J. Math. Phys. \textbf{42}, 2929-2960 (2001)
[arXiv:hep-th/0001053 [hep-th]].


\bibitem{Maldacena:2001km}
J.~M.~Maldacena and H.~Ooguri,
{\em Strings in AdS(3) and the SL($2, \mathbb{R}$) WZW model. Part 3. Correlation functions,}
Phys. Rev. D \textbf{65}, 106006 (2002)
[arXiv:hep-th/0111180 [hep-th]].


\bibitem{Seiberg:1990eb}
N.~Seiberg,
{\em Notes on quantum Liouville theory and quantum gravity,}
Prog. Theor. Phys. Suppl. \textbf{102}, 319-349 (1990)



\bibitem{Gurarie:1993xq}
V.~Gurarie,
{\em Logarithmic operators in conformal field theory,}
Nucl. Phys. B \textbf{410}, 535-549 (1993)
doi:10.1016/0550-3213(93)90528-W
[arXiv:hep-th/9303160 [hep-th]].


\bibitem{Nagi:2005cm}
J.~Nagi,
{\em Logarithmic primary fields in conformal and superconformal field theory,}
Nucl. Phys. B \textbf{722}, 249-265 (2005)
doi:10.1016/j.nuclphysb.2005.05.007
[arXiv:hep-th/0504009 [hep-th]].


\bibitem{Nivesvivat:2020gdj}
R.~Nivesvivat and S.~Ribault,
{\em Logarithmic CFT at generic central charge: from Liouville theory to the $Q$-state Potts model,}
SciPost Phys. \textbf{10}, 021 (2021)
doi:10.21468/SciPostPhys.10.1.021
[arXiv:2007.04190 [hep-th]].



\bibitem{Kutasov:1999xu}
D.~Kutasov and N.~Seiberg,
{\em More comments on string theory on AdS(3),}
JHEP \textbf{04}, 008 (1999)
[arXiv:hep-th/9903219 [hep-th]].


\bibitem{Teschner:2001rv}
J.~Teschner,
{\em Liouville theory revisited,}
Class. Quant. Grav. \textbf{18}, R153-R222 (2001)
[arXiv:hep-th/0104158 [hep-th]].



\bibitem{Ginsparg:1993is}
P.~H.~Ginsparg and G.~W.~Moore,
{\em Lectures on 2-D gravity and 2-D string theory,}
[arXiv:hep-th/9304011 [hep-th]].



\bibitem{DiFrancesco:1997nk}
P.~Di Francesco, P.~Mathieu and D.~Senechal,
{\em Conformal Field Theory,}
doi:10.1007/978-1-4612-2256-9



\bibitem{Maldacena:2000kv}
J.~M.~Maldacena, H.~Ooguri and J.~Son,
{\em Strings in AdS(3) and the SL($2, \mathbb{R}$) WZW model. Part 2. Euclidean black hole,}
J. Math. Phys. \textbf{42}, 2961-2977 (2001)
[arXiv:hep-th/0005183 [hep-th]].


\bibitem{Erbin}
H.~Erbin,
{\em Notes on 2d quantum gravity and Liouville theory},
\url{https://www.lpthe.jussieu.fr/~erbin/files/liouville_theory.pdf},
2015



\bibitem{Harlow:2011ny}
D.~Harlow, J.~Maltz and E.~Witten,
{\em Analytic Continuation of Liouville Theory,}
JHEP \textbf{12}, 071 (2011)
[arXiv:1108.4417 [hep-th]].



\bibitem{Ribault:2005wp}
S.~Ribault and J.~Teschner,
{\em H+(3)-WZNW correlation functions from Liouville theory,}
JHEP \textbf{06}, 014 (2005)
[arXiv:hep-th/0502048 [hep-th]].


\bibitem{Chang:2014jta}
C.~M.~Chang, Y.~H.~Lin, S.~H.~Shao, Y.~Wang and X.~Yin,
{\em Little String Amplitudes (and the Unreasonable Effectiveness of 6D SYM),}
JHEP \textbf{12}, 176 (2014)
[arXiv:1407.7511 [hep-th]].


\bibitem{Iguri:2007af}
S.~Iguri and C.~A.~Nunez,
{\em Coulomb integrals for the SL($2, \mathbb{R}$) WZW model,}
Phys. Rev. D \textbf{77}, 066015 (2008)
[arXiv:0705.4461 [hep-th]].


\bibitem{Teschner:1995yf}
J.~Teschner,
{\em On the Liouville three point function,}
Phys. Lett. B \textbf{363}, 65-70 (1995)
[arXiv:hep-th/9507109 [hep-th]].



\bibitem{Kupiainen:2018snr}
A.~Kupiainen, R.~Rhodes and V.~Vargas,
{\em The DOZZ Formula from the Path Integral,}
JHEP \textbf{05}, 094 (2018)
[arXiv:1803.05418 [hep-th]].

\bibitem{GSW}
M. B. Green, J. H. Schwarz, and E. Witten,
{\em Superstring Theory: 1,}
Cambridge University Press, 1987.


\end{thebibliography}
\end{document}